%% file: VQ4Rec.tex
\documentclass[sigconf]{acmart}
\AtBeginDocument{%
  \providecommand\BibTeX{{%
    \normalfont B\kern-0.5em{\scshape i\kern-0.25em b}\kern-0.8em\TeX}}}

\setcopyright{acmlicensed}
\copyrightyear{2018}
\acmYear{2018}
\acmDOI{XXXXXXX.XXXXXXX}

\acmConference[Conference acronym 'XX]{Make sure to enter the correct
  conference title from your rights confirmation emai}{June 03--05,
  2018}{Woodstock, NY}
\acmISBN{978-1-4503-XXXX-X/18/06}

\usepackage{multirow}
\usepackage{colortbl}
\usepackage{pgfplots}
\usepackage{makecell}
\usepackage{fontawesome}
\usepackage{xcolor}

\definecolor{cvq}{HTML}{3f793a}
\definecolor{cpq}{HTML}{465581}
\definecolor{csq}{HTML}{7E3838}
\newcommand{\tvq}[1]{\textcolor{cvq}{\texttt{#1}}}
\newcommand{\tpq}[1]{\textcolor{cpq}{\texttt{#1}}}
\newcommand{\tsq}[1]{\textcolor{csq}{\texttt{#1}}}

\begin{document}

\title[Vector Quantization for Recommender Systems: A Review and Outlook]{Vector Quantization for Recommender Systems: \\A Review and Outlook}

\input{Content/Author}

\renewcommand{\shortauthors}{Liu, et al.}

\input{Content/Abstract}

\input{Content/CCS}


\maketitle

\input{Content/Introduction}

\input{Content/Concepts}
\input{Content/Taxonomy}
\input{Content/Efficiency}
\input{Content/Quality}
\input{Content/Challenge}
\input{Content/Conclusion}
\input{Content/Acknowledgement}

\bibliographystyle{ACM-Reference-Format}
\bibliography{VQ4Rec}

\end{document}

%% file: Content/Author.tex
\author{Qijiong Liu}
\authornote{\hspace{0.5ex}Equal contribution.}
\affiliation{%
  \institution{The HK PolyU}
  \city{Hong Kong SAR}
  \country{China}
}
\email{liu@qijiong.work}

\author{Xiaoyu Dong}
\authornotemark[1]
\affiliation{%
  \institution{The HK PolyU}
  \city{Hong Kong SAR}
  \country{China}
}
\email{dongxiaoyu7313@gmail.com}

\author{Jiaren Xiao}
\affiliation{%
  \institution{The HK PolyU}
  \city{Hong Kong SAR}
  \country{China}
}
\email{jiaren.xiao@polyu.edu.hk}

\author{Nuo Chen}
\affiliation{%
  \institution{The HK PolyU}
  \city{Hong Kong SAR}
  \country{China}
}
\email{napnap.chen@connect.polyu.hk}

\author{Hengchang Hu}
\affiliation{%
  \institution{National University of Singapore}
  \city{}
  \country{Singapore}
}
\email{hengchang.hu@u.nus.edu}

\author{Jieming Zhu}
\affiliation{%
  \institution{Huawei Noah's Ark Lab}
  \city{Shenzhen}
  \country{China}
}
\email{jiemingzhu@ieee.org}

\author{Chenxu Zhu}
\affiliation{%
  \institution{Huawei Noah's Ark Lab}
  \city{Shanghai}
  \country{China}
}
\email{zhuchenxu1@huawei.com}

\author{Tetsuya Sakai}
\affiliation{%
  \institution{Waseda University}
  \city{Tokyo}
  \country{Japan}
}
\email{tetsuyasakai@acm.org}

\author{Xiao-Ming Wu}
\authornote{Corresponding author.}
\affiliation{
  \institution{The HK PolyU}
  \city{Hong Kong SAR}
  \country{China}
}
\email{xiao-ming.wu@polyu.edu.hk}

%% file: Content/Abstract.tex
\begin{abstract}
Vector quantization, renowned for its unparalleled feature compression capabilities, has been a prominent topic in signal processing and machine learning research for several decades and remains widely utilized today. With the emergence of large models and generative AI, vector quantization has gained popularity in recommender systems, establishing itself as a preferred solution. This paper starts with a comprehensive review of vector quantization techniques. It then explores systematic taxonomies of vector quantization methods for recommender systems (VQ4Rec), examining their applications from multiple perspectives. Further, it provides a thorough introduction to research efforts in diverse recommendation scenarios, including efficiency-oriented approaches and quality-oriented approaches. Finally, the survey analyzes the remaining challenges and anticipates future trends in VQ4Rec, including the challenges associated with the training of vector quantization, the opportunities presented by large language models, and emerging trends in multimodal recommender systems. We hope this survey can pave the way for future researchers in the recommendation community and accelerate their exploration in this promising field.
\end{abstract}

%% file: Content/CCS.tex
\ccsdesc[500]{Information systems~Recommender systems}
\ccsdesc[500]{General and reference~Surveys and overviews}

%% file: Content/Introduction.tex
\section{Introduction}


Vector quantization~\cite{buzo1980speech,vq} (VQ), a cornerstone technique in signal processing, was originally introduced by Gray and his team~\cite{buzo1980speech} in the 1980s to compress data representation while preserving the fidelity of the original signal. The foundational standard VQ technique aims to compress the entire representation space into a compact codebook containing multiple codewords, typically using a single code to approximate each vector. To improve the precision of quantization, advanced methods such as product quantization~\cite{sabin1984product} and residual quantization~\cite{juang1982multiple,gray1998quantization,rq} were introduced, representing {parallel} and {sequential} approaches, respectively. These VQ techniques have proven to be highly effective in domains including speech~\cite{makhoul1985vector,abe1990voice} and image coding~\cite{nasrabadi1988image,cosman1993using}.


\input{Illustration/Figures/paper_count}

Despite its early development, it was not until the late 1990s that VQ found application in the field of information retrieval, particularly in image retrieval~\cite{lu1999novel}. The progress in applying VQ techniques was slow until 2010 when Jegou and his team~\cite{pq} demonstrated the effectiveness of parallel quantization for approximate nearest neighbor search. This innovation enables fast similarity computations in high-dimensional data spaces. In the same year, Chen and his team~\cite{chen2010approximate} investigated the potential of sequential quantization for similar applications. 

\input{Illustration/Figures/vqs}

Recommender systems, a prominent application in the field of artificial intelligence and data science, typically build upon advancements in information retrieval and machine learning. The integration of VQ into recommender systems started in 2004, initially applied to music recommendation~\cite{huang2004audio}. However, a major turning point occurred 15 years later, sparked by the introduction of VQ-VAE~\cite{vq-vae} for image generation, which utilized VQ to discretize image representations. This innovation led to the development of PQ-VAE~\cite{van2019pq}, which brought renewed attention to VQ within the recommendation community. The success of VQ-VAE also catalyzed further advancements in residual quantization, leading to the creation of RQ-VAE~\cite{rq-vae}, which is now at the heart of the burgeoning field of generative recommender systems~\cite{rajput2023recommender}. Furthermore, the emergence of large language models (LLMs)~\cite{gpt-4,llama-2} has spurred new applications in the recommendation domain~\cite{liu2024once}. However, due to their substantial size and latency during inference, there's a growing trend in recommender systems to adopt VQ to enhance efficiency.

As shown in Figure~\ref{fig:papercount}, there has been a booming interest in vector quantization for recommender systems (\textbf{VQ4Rec}) over recent years.
This body of research can be roughly categorized into \textit{\textbf{efficiency-oriented}} and \textit{\textbf{quality-oriented}}.
The former focuses on optimizing large-scale systems, tackling challenges associated with large models, extensive datasets, and computational demands. In this context, VQ proves to be highly effective, significantly improving performance in crucial areas, including \textbf{similarity search}~\cite{su2023beyond}, \textbf{space compression}~\cite{imran2023refrs}, and \textbf{model acceleration}~\cite{wu2021linear}. The latter prioritizes recommendation accuracy, concentrating on the refinement of feature usage. This involves optimizing features, fostering interactions among various modalities, and aligning features to enhance generative recommendation processes. It covers sub-scenarios such as \textbf{feature enhancement}~\cite{luo2024within}, \textbf{modality alignment}~\cite{hou2023learning}, and \textbf{discrete tokenization}~\cite{rajput2023recommender}. 
Moreover, VQ has shown promise in integrating recommender systems with LLMs~\cite{zheng2023adapting} to improve recommendation quality. 
This is achieved by using VQ to effectively tokenize and structure recommendation-related data, such as information about items or users. For instance, generative retrieval methods~\cite{rajput2023recommender} leverage VQ to ensure that the recommendation data is well-aligned with LLMs. 

Despite the growing interest in VQ4Rec amidst new challenges posed by large language models, multimodal data, and generative AI, no work has yet systematically surveyed the application of VQ in recommender systems. This paper aims to bridge this gap through a comprehensive survey. We provide a thorough analysis of VQ4Rec, exploring its uses, challenges, and future directions in the field. The main contents and contributions of this paper are summarized as follows:
\begin{itemize}
    \item We present an overview of both classical and modern VQ techniques, encompassing standard VQ, parallel VQ, sequential VQ, and differentiable VQ.
    \item We provide systematic taxonomies of VQ4Rec from various perspectives such as training phase, application scenario, VQ techniques, and quantization target.
    \item We conduct a thorough analysis of the strengths, weaknesses, and limitations of existing VQ4Rec methods, focusing on addressing two main challenges in recommender system: efficiency and quality.
    \item We identify key challenges in VQ4Rec and present promising opportunities that can serve as inspiration for future research in this burgeoning field.
\end{itemize}

%% file: Illustration/Figures/paper_count.tex
\begin{figure}[t]
    \centering
    \includegraphics[width=.9\linewidth]{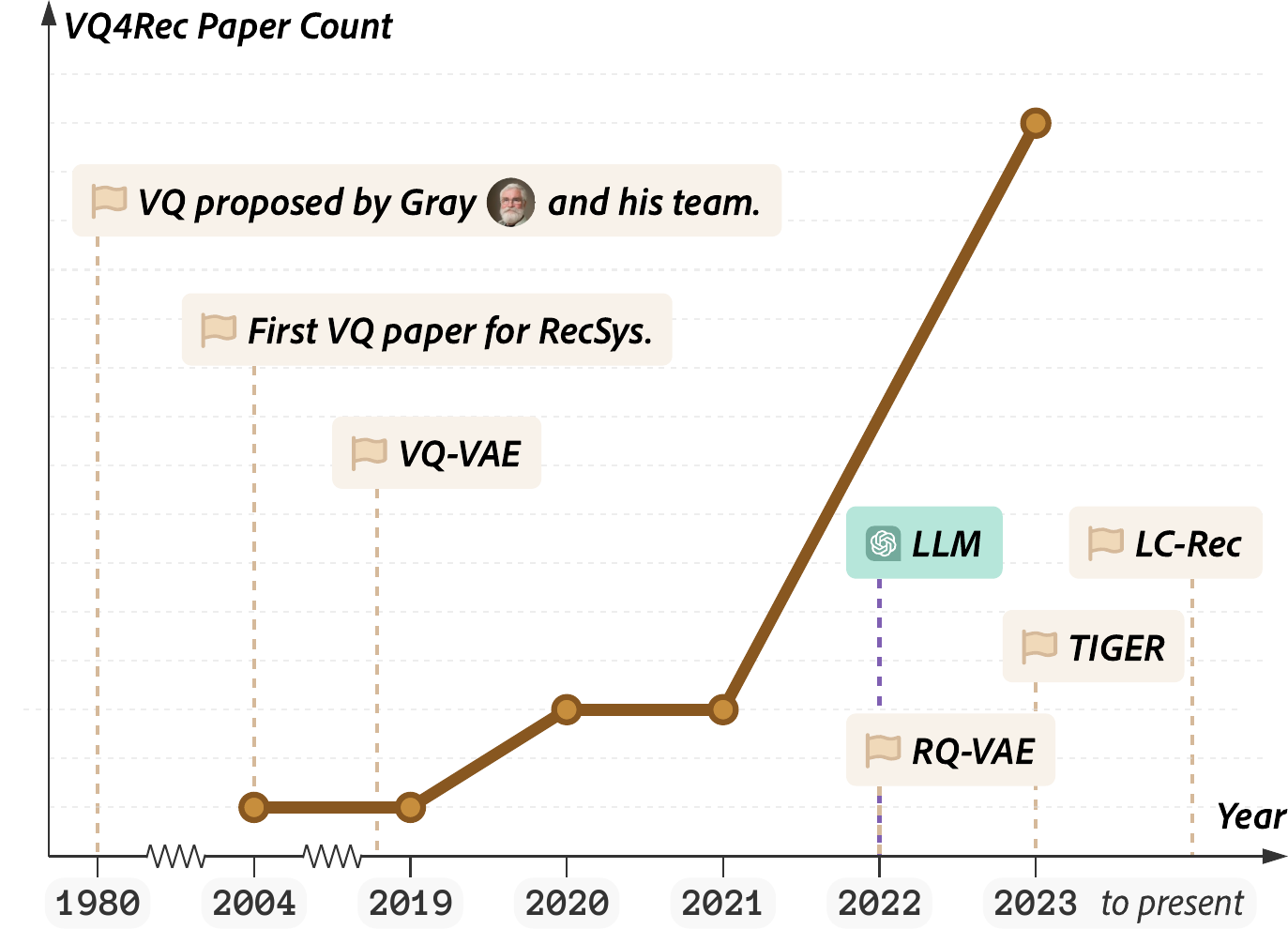}
    \caption{Interest in VQ4Rec over time. \faFlag{} denotes a milestone event or a representative paper.}
    \label{fig:papercount} 
\end{figure}

%% file: Illustration/Figures/vqs.tex
\begin{figure*}[ht]
    \centering
    \includegraphics[width=.9\linewidth]{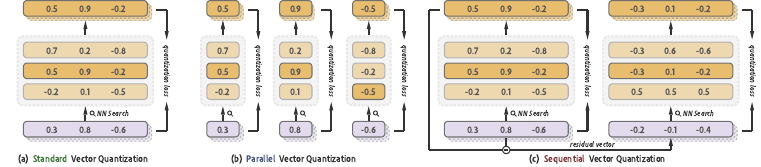}
    \caption{Illustration of the three classical VQ techniques. \faSearch{} indicates nearest neighbor search.}
    \label{fig:vqs} 
\end{figure*}

%% file: Content/Concepts.tex
\input{Illustration/Tables/quantizer}

\section{Overview of VQ Techniques }

VQ targets at grouping similar vectors into clusters by representing them with a small set of prototype vectors (i.e., codes in the codebook). In this section, we offer a comprehensive summary of classical VQ methods and the modern differentiable VQ technique. The conventional VQ approaches include standard VQ, which uses a single codebook, parallel VQ, which utilizes multiple codebooks simultaneously to represent separate vector subspaces, and sequential VQ, which involves using multiple codebooks in a sequence to refine the quantization.



\subsection{Standard Vector Quantization}

The standard VQ~\citep{buzo1980speech,vq} serves as the atomic component for the latter two VQ techniques. Formally, given a set of object vectors $\mathbf{E} \in \mathbb{R}^{N \times D}$, a function $f$ (e.g., $k$-means) is required to produce a codebook $\mathbf{C} \in \mathbb{R}^{K \times D}$ such that the sum of distances between all vectors in $\mathbf{E}$ and their corresponding nearest code vectors in $\mathbf{C}$ is minimized, as illustrated in Figure~\ref{fig:vqs}(a). We can formally express this using the following equations:
\begin{align}
    f&: \mathbf{E} \rightarrow \mathbf{C}, \\
    \textit{where }\mathbf{C} &= \underset{\mathbf{W} \in \mathbb{R}^{K \times D}}{\operatorname{argmin}} \sum_{i=1}^{N} d(\mathbf{e}_i, \mathbf{w}_{x}), \\
    \textit{and }x &= \underset{j=1,\ldots,K}{\operatorname{argmin}}\, d\left(\mathbf{e}_i, \mathbf{w}_j\right),
\end{align}
where $N$ is the number of object vectors and $K$ is the number of code vectors in the codebook (usually $N \gg K$), $\mathbf{e}_i$ is the $i$-th object vector, $D$ is the embedding dimension, $d$ represents the distance function (e.g., Euclidean distance or Manhattan distance), $\mathbf{W}$ denotes any codebook in the same space as $\mathbf{C}$, and $x$ is the index of the code vector closest to $\mathbf{e}_i$. Therefore, we can use $\mathbf{c}_x$, the $x$-th code in codebook $\mathbf{C}$, to approximate $\mathbf{e}_i$:
\begin{equation}
    \mathbf{e}_i \approx \mathbf{c}_{x}.
\end{equation}

\subsection{Parallel Vector Quantization}

As the embedding dimension $D$ increases, standard VQ methods face significant challenges in terms of storage requirements, computational efficiency, and quantization quality. In response to these challenges, approaches like product quantization and optimized product quantization, representative of parallel quantization techniques, emerge as effective solutions.
These methods segment high-dimensional vectors into multiple lower-dimensional sub-vectors and perform quantization on each segment independently. As shown in Table~\ref{tab:quantizers}, with an increase in the number of segments ($M$), there is a corresponding reduction in the dimensionality of each code, keeping the codebook storage size unchanged. Yet, the representation space exhibits an exponential growth compared to that of standard VQ.

\subsubsection{Product Quantization (PQ)~\citep{juang1982multiple,pq}}\label{sec:pq} Product Quantization (PQ) represents an initial approach to parallel quantization, where original high-dimensional vectors are segmented into uniformly-sized sub-vectors. This process can be mathematically represented as
$\mathbf{E} = \left[\mathbf{E}^{(1)}, \mathbf{E}^{(2)}, \cdots, \mathbf{E}^{(M)} \right]$,
where $M$ denotes the number of the segments and the number of the codebooks, and $\mathbf{E}^{(i)} \in \mathrm{R}^{N \times \frac{D}{M}}$. Each sub-vector is then independently subjected to standard VQ, utilizing a distinct codebook for each segment.
Therefore, the $i$-th original vector can be approximated by selecting and concatenating each single code vector $\mathbf{c}^{(j)}_{x_j}$ from each sub-codebook $\mathbf{C}^{(j)}$, which can be formulated as:
\begin{equation}
    \mathbf{e}_i = \left[\cdots, \mathbf{e}_i^{(j)}, \cdots \right] \approx \left[\cdots, \mathbf{c}^{(j)}_{x_j}, \cdots,\right] \quad \text{for } j \in \{1, 2, \ldots, M\},
\end{equation}
where $\mathbf{C}^{(j)}$ is the $j$-th codebook with size $K_j$, and $x_j$ is the index of the code vector in $\mathbf{C}^{(j)}$ closest to $\mathbf{e}_i^{(j)}$.
Due to its storage efficiency and capability for fast approximate nearest neighbor searches, product quantization has become a popular solution in the information retrieval domain, particularly for image retrieval tasks, as evidenced by several studies~\cite{cao2017deep,jang2021self,chen2022adversarial}. Nonetheless, it overlooks the potential for significant inter-correlations among sub-vectors, which may affect the quantization performance and subsequent downstream tasks.

\subsubsection{Optimized Product Quantization (OPQ)~\citep{opq}} To eliminate the interdependence among multiple subspaces, optimized product quantization is introduced and uses the learnable rotation matrix $\mathbf{R} \in \mathbb{R}^{D \times D}$ for automatically selecting the most effective orientation of the data in the high-dimensional space. Such rotation minimizes the interdependence among different subspaces, allowing for a more efficient and independent quantization process, which can be defined as:
\begin{align}
    \mathbf{E}^\prime &= \mathbf{E} \times \mathbf{R}, \\
    \mathbf{I} &= \mathbf{R}^T \times \mathbf{R},
\end{align}
where $\mathbf{E}^\prime$ is the rotated matrix, and $\mathbf{I}$ represents the identity matrix. Next, $\mathbf{E}^\prime$ will be operated by product quantization, as described in Sec~\ref{sec:pq}. It is important to note that the rotation matrix $\mathbf{R}$ is trained with the codebooks.
Once trained, the $i$-th original vector can be approximated by:
\begin{equation}
    \mathbf{e}_i \approx \left[\cdots, \mathbf{c}^{(j)}_{x_j}, \cdots,\right] \times \mathbf{R}^T \quad \text{for } j \in \{1, 2, \ldots, M\}.
\end{equation}

\subsection{Sequential Vector Quantization}

Standard VQ and parallel VQ typically yield \emph{rough} approximations of vectors. Specifically, each dimension of the original vector can only be approximated by one single value from the corresponding code vector, leading to substantial information loss. 
Taking standard VQ as an example, the difference between the original vector $\mathbf{e}$ and its corresponding code $\mathbf{c}$, denoted by $\mathbf{e}-\mathbf{c}$, reflects the unique characteristics that cannot be represented by $\mathbf{c}$.

To achieve a more \emph{precise} quantization, approaches like residual quantization~\cite{juang1982multiple,rq} and additive quantization~\cite{aq} have been developed, falling under the umbrella of sequential quantization. This method employs multiple codebooks, with each codebook approximates every dimension of the original vectors. Essentially, every codebook offers a distinct approximation perspective of the vectors, and the accuracy of these approximations improves with an increase in the number of codebooks. As illustrated in Figure~\ref{fig:vqs}, using the first layer codebook approximates \texttt{0.3} (the first dimension of the original vector) as \texttt{0.5} (the first dimension of the code vector in the first codebook). After applying the second codebook, it is more accurately approximated as 0.5 + (-0.3) = 0.2 (the first dimension of the code vector in the second codebook).

\subsubsection{Residual Quantization (RQ)~\citep{juang1982multiple,rq}} By designing $M$ individual codebooks where, as depicted in Table~\ref{tab:quantizers}, code vectors have the full same length of the input vector, residual quantization aims to approximate the target vectors by compressing their information in a coarse-to-fine manner. Specifically, the codebooks are learned iteratively from the residual representations of the vectors. This process can be formulated as: $\mathbf{E}^{(j+1)} = \mathbf{E}^{(j)} - \mathbf{X}^{(j)}\mathbf{C}^{(j)}$, where $\mathbf{E}^1 = \mathbf{E}$, $\mathbf{C}^{(j)}$ is the $j$-th codebook with size $K_j$, and $\mathbf{X}^{(j)} \in \mathrm{\{0, 1\}}^{N}$ is a one-hot mapper, where $\mathbf{X}^{(j)}_{i,k}=1$ only if the $k$-th code is the nearest to the $i$-th vector of $\mathbf{E}^{(j)}$ in the codebook $\mathbf{C}^{(j)}$. After iteratively residual approximation, the $i$-th original vector can be represented by:
\begin{align}
    \mathbf{e}_i &\approx \sum_j^M \mathbf{c}^{(j)}_{x_j}, \label{eq:sequential} \\
    \textit{where }x_j &= \underset{k}{\operatorname{argmin}}\,\mathbf{X}^{(j)}_{i,k}.
\end{align}
It is important to note that, as $M$ increases, the approximated representation tends to be finer.

\subsubsection{Additive Quantization (AQ)~\citep{aq}} Similar to residual quantization, additive quantization aims to approximate the target vectors by aggregating one selected code per codebook. However, residual quantization employs a greedy approach by selecting only the \emph{nearest} neighbor (i.e., $\mathbf{c}^{(j)}_{x_j}$) within the current (i.e., $j$-th) layer, which does not guarantee the global optimum. Instead, codebooks here are sequentially learned using beam search, where top candidate code combinations (\emph{not the only one}) from the first $j$ codebooks are selected to infer the $(j+1)$-th codebook. Hence, the $i$-th original vector can be approximated as in Equation~\ref{eq:sequential}.

\subsection{Differentiable Vector Quantization}

The technique of VQ fundamentally includes a non-differentiable procedure, which entails identifying the nearest code in the codebook, consequently making the calculation of gradients impractical. This lack of differentiability presents a substantial hurdle in neural network training, which relies heavily on gradient-based optimization methods. Consequently, in the wake of the VQ-VAE~\cite{vq-vae}, numerous research initiatives~\citep{kang2020learning,rajput2023recommender} have adopted the Straight-Through Estimator (STE)~\citep{ste} as a leading solution to this challenge.

The core idea of STE is relatively straightforward: during the forward pass of a network, the non-differentiable operation (like quantization) is performed as usual. However, during the backward pass, when gradients are propagated back through the network, STE allows gradients to ``pass through'' the non-differentiable operation as if it were differentiable. This is typically done by approximating the derivative of the non-differentiable operation with a constant value, often 1, which can be defined as:
\begin{equation}
    \frac{\partial \mathbf{c}_x}{\partial \mathbf{e}_i} \approx \frac{\partial \mathbf{e}_i}{\partial \mathbf{e}_i} = \mathbf{I},
\end{equation}
where $\mathbf{I}$ is the identity matrix.


However, training with straight-through estimator often encounters the codebook collapse issue, wherein a significant portion of codes fails to map onto corresponding vectors. Various strategies, such as employing exponential moving average (EMA)~\cite{ema} during training or implementing codebook reset~\cite{zeghidour2021soundstream,rajput2023recommender} mechanisms, have been developed to address this challenge.







In the above discussion, we have reviewed established vector quantization techniques, but have not delved into recent innovations such as finite scalar quantization (FSQ)~\cite{donahue2019piano,dieleman2021variable,mentzer2023finite}. Drawing inspiration from model quantization~\cite{shi2023quantize,chen2023hessian,yue2023llamarec}, FSQ adopts a straightforward rounding mechanism to approximate the value in each dimension of a vector. FSQ has yielded competitive results comparable to those achieved by VQ-VAE~\cite{mentzer2023finite} in image generation. While FSQ has not yet been applied to recommender systems, it presents a promising avenue for future exploration.


%% file: Illustration/Tables/quantizer.tex
\begin{table*}[t]
\centering
\renewcommand{\arraystretch}{1.2} 

\caption{Comparison of the three classical VQ techniques. We use $\bar{K}=\frac{1}{M} \sum_i K_i$ to represent the arithmetic mean of $K_i$, and $\hat{K}=\sqrt[M]{\prod_i K_i}$ to represent their geometric mean, where $i \in \{1, 2, \ldots, M\}$. Note that when $K_i=K$, $\bar{K} = \hat{K} = K$.}
\label{tab:quantizers}

\begin{tabular}{c|c|c|c|c|c|c}
\toprule
 & \textbf{Input Dim} & \textbf{\#Codebooks} & \textbf{\#Codes per Book} & \textbf{Code Dim} & \textbf{Codebook Size} & \textbf{Feature Space} \\
\midrule
\textbf{Standard VQ} & $D$ & $1$ & $K$ & $D$ & $K \cdot D$ & $K$ \\
\midrule
\textbf{Parallel VQ} & $D$ & $M$ & $K_i$ & $D / M$ & $\bar{K} \cdot D$ & $\hat{K}^M$ \\
\midrule
\textbf{Sequential VQ} & $D$ & $M$ & $K_i$ & $D$ & $M \cdot \bar{K} \cdot D$ & $\hat{K}^M$ \\
\bottomrule
\end{tabular}

\end{table*}

%% file: Content/Taxonomy.tex
\input{Illustration/Figures/paradigm}

\section{Taxonomies of VQ4Rec}

\input{Illustration/Tables/paperlist}

To comprehensively understand the current advances in VQ4Rec, in this section, we categorize previous studies from multiple viewpoints, such as training phase or application scenario, to encapsulate the diverse methodologies and applications in this field.

\subsection{Classification by Training Phase}\label{sec:paradigm}

VQ techniques can be applied to recommender systems at different training stages: pre-processing, in-processing, and post-processing, as depicted in Figure~\ref{fig:paradigm}. 
\begin{itemize}
    \item \textbf{Pre-processing:} In this stage, VQ techniques are utilized to optimize or compress input data, such as item features or user sequences, resulting in static quantized inputs for recommender systems~\citep{zhao2021embedding,imran2023refrs,hou2023learning}.
    \item \textbf{In-processing:} Here, VQ is integrated to and trained together with the recommender system, providing dynamically quantized features to enhance the functionality of the system~\citep{kang2020learning,van2019pq,wu2021linear}.
    \item \textbf{Post-processing:} This involves applying VQ to the embeddings generated by the recommender systems, aiming to improve search speed or accuracy~\citep{zhang2023query,lu2023differentiable,huang2004audio}.
\end{itemize}

\subsection{Classification by Application Scenario}\label{sec:scenario}

The use of VQ in recommender systems can be broadly classified into two major scenarios: one that prioritizes efficiency and another that emphasizes quality. As depicted in Figure~\ref{fig:applications}, each scenario addresses distinct challenges and objectives inherent to the recommender system, leveraging the strengths of VQ to enhance the overall performance and user experience.

\textbf{Efficiency-oriented approaches} primarily focus on enhancing the computational and storage aspects of recommender systems. In this fast-evolving digital era, where data volume and complexity are ever-increasing, these approaches play a instrumental role in maintaining the scalability and responsiveness of recommendation services. They are particularly pertinent in scenarios such as similarity search~\citep{zhang2023query,lu2023differentiable,zhao2021embedding}, space compression~\citep{imran2023refrs,kang2020learning,van2019pq}, and time acceleration~\citep{wu2021linear,lian2020lightrec,lian2020product}.

Conversely, \textbf{quality-oriented approaches} aim to enhance the accuracy and relevance of the recommendations. These methods leverage VQ to refine the data and model representations, thereby improving the quality of the output provided to the end-users. They are relevant in scenarios involving feature enhancement~\citep{liu2024cage,pan2021click}, modality correlation~\citep{hou2023learning}, and item tokenization~\citep{razavi2019generating,singh2023better,jin2023language,zheng2023adapting}.

\subsection{Other Classification Frameworks}\label{sec:others}

Here, we expand our perspective to explore additional classification frameworks for VQ4Rec. This includes:

\begin{itemize}
    \item \textbf{Classification by VQ Technique}: As previously mentioned, existing studies generally adopt three types of VQ techniques: \textbf{Standard VQ}, as seen in works like~\cite{huang2004audio,imran2023refrs,wu2021linear}, \textbf{Parallel VQ}, featured in studies~\cite{zhang2023query,lu2023differentiable,zhao2021embedding}, and \textbf{Sequential VQ}, highlighted in references~\cite{lian2020lightrec,rajput2023recommender,singh2023better}.
    \item \textbf{Classification by Quantization Target:} The majority of existing research has focused on \textbf{Item Quantization}~\cite{huang2004audio,imran2023refrs,wu2021linear}. This is likely because item features are usually static, whereas user preferences are dynamic. Additionally, the need to compress extensive item datasets due to their large scale and rich content has been a driving factor. Nonetheless, there is also some research on \textbf{User Quantization}~\cite{van2019pq,pan2021click}, as well as studies that investigate both \textbf{Item \& User Quantization}~\cite{lian2020product,liu2024cage}.
\end{itemize}

%% file: Illustration/Figures/paradigm.tex
\begin{figure}[t]
    \centering
    \includegraphics[width=.9\linewidth]{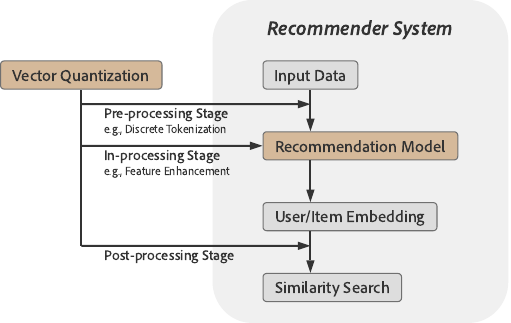}
    \caption{Integration of VQ techniques with the recommender system at different training stages.
    }
    \label{fig:paradigm} 
\end{figure}

%% file: Illustration/Tables/paperlist.tex
\begin{table*}[]
\centering
\renewcommand{\arraystretch}{1.2} 
\setlength\tabcolsep{4pt}
\caption{A list of representative VQ4Rec methods and their features. 
``Modality'' denotes the type of feature utilized, and ``Task'' refers to the specific training tasks employed in these methods. Note that all papers pertaining to the post-processing stage are task-agnostic, hence the ``-'' in the table for these entries. We use ``CTR'', ``NIP'', ``CF'', and ``Multi'' to denote ``click-through rate prediction'', ``next item prediction'', ``collaborative filtering'' and ``multiple'' tasks, respectively.}
\label{tab:all}

\resizebox{\linewidth}{!}{
\begin{tabular}{c|l|l|l|l|l|l|l}
\toprule
\textbf{Application} & \textbf{Paper} & \textbf{Venue} & \textbf{VQ Type} & \textbf{VQ Target} & \textbf{Modality} & \textbf{Stage} & \textbf{Task} \\
\midrule
\multicolumn{8}{c}{\cellcolor[HTML]{F4EFC2} \textbf{Efficiency-Oriented}} \\
\midrule
\multirow{4}{*}{\makecell{Space \\ Compression}}
& \citet{liu2024semantic} & TheWebConf (2024) & \tsq{Sequential / RQ} & Item \& User & ID \& Text & Pre & CTR \\
& \citet{imran2023refrs} & TOIS (2023) & \tvq{Standard VQ} & Item & ID & Pre & NIP \\
& \citet{kang2020learning} & TheWebConf (2020) & \tpq{Parallel / PQ} & Item & ID & In & Multi \\
& \citet{van2019pq} & RecSys (2019) & \tpq{Parallel / PQ} & User & ID & In & CF \\
\midrule
Model Acc.
& \citet{wu2021linear} & TheWebConf (2021) & \tvq{Standard VQ} & Item & ID & In & NIP\\
\midrule
\multirow{7}{*}{\makecell{Similarity \\ Search}}
& \citet{su2023beyond} & SIGIR (2023) & \tvq{Parallel / PQ} & User & ID & Post & - \\
& \citet{zhang2023query} & AAAI (2023) &  \tpq{Parallel / PQ} & Item & ID \& Text & Post & - \\
& \citet{lu2023differentiable} & TheWebConf (2023) & \tpq{Parallel / OPQ} & Item & ID & Post & - \\
& \citet{zhao2021embedding} & KDD-IRS (2021) & \tpq{Parallel / OPQ} & Item & Text & Pre & CTR \\
& \citet{lian2020product} & TKDE (2020) & \tpq{Parallel / OPQ} & Item \& User & ID & In & CF \\
& \citet{lian2020lightrec} & TheWebConf (2020) & \tsq{Sequential / RQ} & Item & ID \& Text & In & CF \\
& \citet{huang2004audio} & ICME (2004) & \tvq{Standard VQ} & Item & Music & Post & - \\
\midrule
\multicolumn{8}{c}{\cellcolor[HTML]{F4EFC2} \textbf{Quality-Oriented}} \\
\midrule
\multirow{3}{*}{\makecell{Feature \\ Enhancement}}
& \citet{liu2024cage} & TheWebConf (2024) & \tvq{Standard VQ} & Item \& User & ID & In & Multi \\
& \citet{luo2024within} & arXiv (2024) & \tvq{Standard VQ} & Item & ID & In & NIP \\
& \citet{pan2021click} & arXiv (2021) & \tvq{Standard VQ} & User & ID & In & CTR \\
\midrule
\multirow{2}{*}{\makecell{Modality \\ Alignment}}
& \citet{hu2024lightweight} & ECIR (2024) & \tpq{Parallel / PQ} & Item & Image \& Text \& ID & In & NIP \\
& \citet{hou2023learning} & TheWebConf (2023) &  \tpq{Parallel / OPQ} & Item & Text & Pre & NIP \\
\midrule
\multirow{6}{*}{\makecell{Discrete \\ Tokenization}}
& \citet{zheng2023adapting} & ICDE (2024) & \tsq{Sequential / RQ} & Item & Text & Pre & NIP \\
& \citet{liu2024mmgrec} & arXiv (2024) & \tsq{Sequential / RQ} & Item \& User & Graph & Pre & CF \\
& \citet{jin2024contrastive} & arXiv (2024) & \tsq{Sequential / RQ} & Item & Text & Pre & NIP \\
& \citet{rajput2023recommender} & NeurIPS (2023) & \tsq{Sequential / RQ} & Item & Text & Pre & NIP \\
& \citet{singh2023better} & arXiv (2023) & \tsq{Sequential / RQ} & Item & Video & Pre & CTR \\
& \citet{jin2023language} & arXiv (2023) & \tvq{Standard VQ} & Item & Text & Pre & NIP \\
\bottomrule
\end{tabular}
}
\end{table*}

%% file: Content/Efficiency.tex
\input{Illustration/Figures/applications}

\section{Efficiency-oriented Approaches}

Efficiency in machine learning is crucial for enhancing model speed and optimizing resource use in environments with limited computational power~\cite{schwartz2020green,liu2024benchmarking}. Advances in technology have led to various solutions to improve model efficiency~\cite{menghani2023efficient}, such as model pruning~\cite{beel2019data}, 
model distillation~\cite{tang2018ranking}, 
and model quantization~\cite{ko2021mascot}.
Moreover, adopting efficient architectures like parameter-efficient finetuning~\cite{liu2024once} or linear attention networks~\cite{liu2023linrec} optimizes training and inference processes without increasing space requirements.


VQ enhances the efficiency of recommender systems with its superior clustering capabilities, being widely used and verified in similarity search, space compression, and model acceleration scenarios.


\subsection{Space Compression}

Recommender systems typically create a unique embedding vector for each user or item, leading to high memory costs with large datasets. For example, 1 billion users would need 238 GB for 64-dimensional vectors in 32-bit floating point~\citep{chen2023clustered}. To mitigate these costs, techniques like hashing~\cite{zhang2018efficient} and low-rank factorization~\cite{koren2009matrix} have been used. However, hashing can cause information loss due to hash collisions, while low-rank factorization might overlook complex data patterns, reducing model accuracy.

One line of research focuses on quantizing and condensing \emph{sequential data}, such as user behavior or item content, using a variational autoencoder mechanism inspired by VQ-VAE~\cite{vq-vae} in image generation. These methods integrate sequential knowledge into a unified representation, subsequently compressed into discrete codes. For example, \citet{van2019pq} introduced PQ-VAE, employing product quantization to derive discrete user representations from user-item interactions for quick prediction of click-through rates. Similarly, ReFRS~\cite{imran2023refrs} uses a variational autoencoder within a federated learning framework to learn user tokens for decentralized recommendations. Recently, \citet{liu2024semantic} introduces residual quantization to condense both user history and item content into short tokens. Compared with embedding-based models, caching these tokens would achieve about 100x space compression rate. Another research approach directly applies VQ to existing \emph{embedding tables}, as exemplified by MGQE~\cite{kang2020learning}, which utilizes differentiable VQ for item embeddings.

These methods often also accelerate training and inference through more streamlined model architectures. However, VQ techniques have yet to be empirically tested for space compression in large-scale recommendation models, where their feasibility may be challenged by high embedding dimensions.

\subsection{Model Acceleration}
Prior section has investigated methods for enhancing training and inference efficiency through space compression and dimensionality reduction. 
Here, we focus on summarizing research aimed at accelerating the model architecture.

Transformers and attention mechanisms~\cite{vaswani2017attention}, fundamental to many influential models, exhibit inference efficiency that scales quadratically with sequence length. Consequently, significant researches have been directed toward developing attention modules that operate with linear time complexity. Techniques such as low-rank matrix decomposition (used in Linformer~\cite{linformer} and Performer~\cite{performer}) and hashing for matching attention values (used in EcoFormer~\cite{ecoformer}) have been explored. Additionally, VQ, which applies clustering to condense the attention matrix space, has demonstrated efficacy in fields like time series forecasting and natural language processing. 
Notably, Wu et al. propose LISA~\cite{wu2021linear} which expedites inference for long-sequence recommender systems. Compared with existing approaches which apply sparse attention patterns where crucial information may be lost, LISA combines the effectiveness of self-attention and the efficiency of sparse attention, enabling full contextual attention through codeword histograms.

Currently, the application of VQ for model optimization and acceleration remains limited. 
However, VQ-based linear attention modules are likely to gain popularity with the increase in long sequence features and the emergence of lifelong learning in the era of big data. Additionally, recent studies have employed VQ for the identification and compression of graph structures, followed by distillation of the compressed features into MLP format~\cite{yang2023vqgraph}. 
This approach enhances the processing of graph structural information, offering potential benefits for graph-based recommender systems, such as in social recommendation contexts.

\subsection{Similarity Search}

Similarity search, which relies on recommendation models for learning user and item representations, enables the retrieval of similar users and items. In 2004, \citet{huang2004audio} first highlighted the robust matching capabilities of VQ for music recommendation, categorizing new music representations into pre-existing groups using nearest neighbor search. However, conducting exhaustive maximum inner product searches (MIPS) is often costly and impractical with a large number of candidates. To mitigate these issues, a substantial body of research has focused on approximate nearest neighbor search (ANNs) and MIPS techniques, including hashing~\cite{neyshabur2015symmetric}, tree search~\cite{feng2023reinforcement}, and graph search~\cite{morozov2018non}. 

In 2010, \citet{pq} pioneered a novel solution in the similarity search domain by employing a divide-and-conquer strategy, which involved subdividing vectors into sub-vectors followed by quantization. This product quantization based method facilitates rapid estimation of approximate distances between vectors represented by codes, through the pre-computation of distance tables for each code. This efficient technique for approximate nearest neighbors (ANNs) quickly became a mainstream approach in similarity search, including \emph{item-item search}~\cite{johnson2019billion,lu2023differentiable,zhang2023query,zhao2021embedding} and \emph{user-item search}~\cite{lian2020product,su2023beyond}. Beyond parallel quantization methods, \citet{lian2020lightrec} have explored sequential quantization to discretize item embeddings, thereby enhancing relevance score estimation and reducing memory requirements in recommender systems.

Parallel and sequential quantization both aim to establish one-to-one mappings between vectors and code combinations, expanding horizontally and vertically, respectively, and have been validated in similarity search. However, there is currently no method that combines these approaches to finely segment and represent vectors. Additionally, similarity search techniques for weight matrices and recent low-rank adaptation (LoRA)~\cite{hu2021lora} methods share similarities in achieving approximate effects through matrix compression. In the future, these methods may also find application in parameter-efficient finetuning~\cite{liu2022few} for recommendations, offering potential new directions for efficiency-oriented applications.

%% file: Illustration/Figures/applications.tex
\begin{figure}[t]
    \centering
    \includegraphics[width=\linewidth]{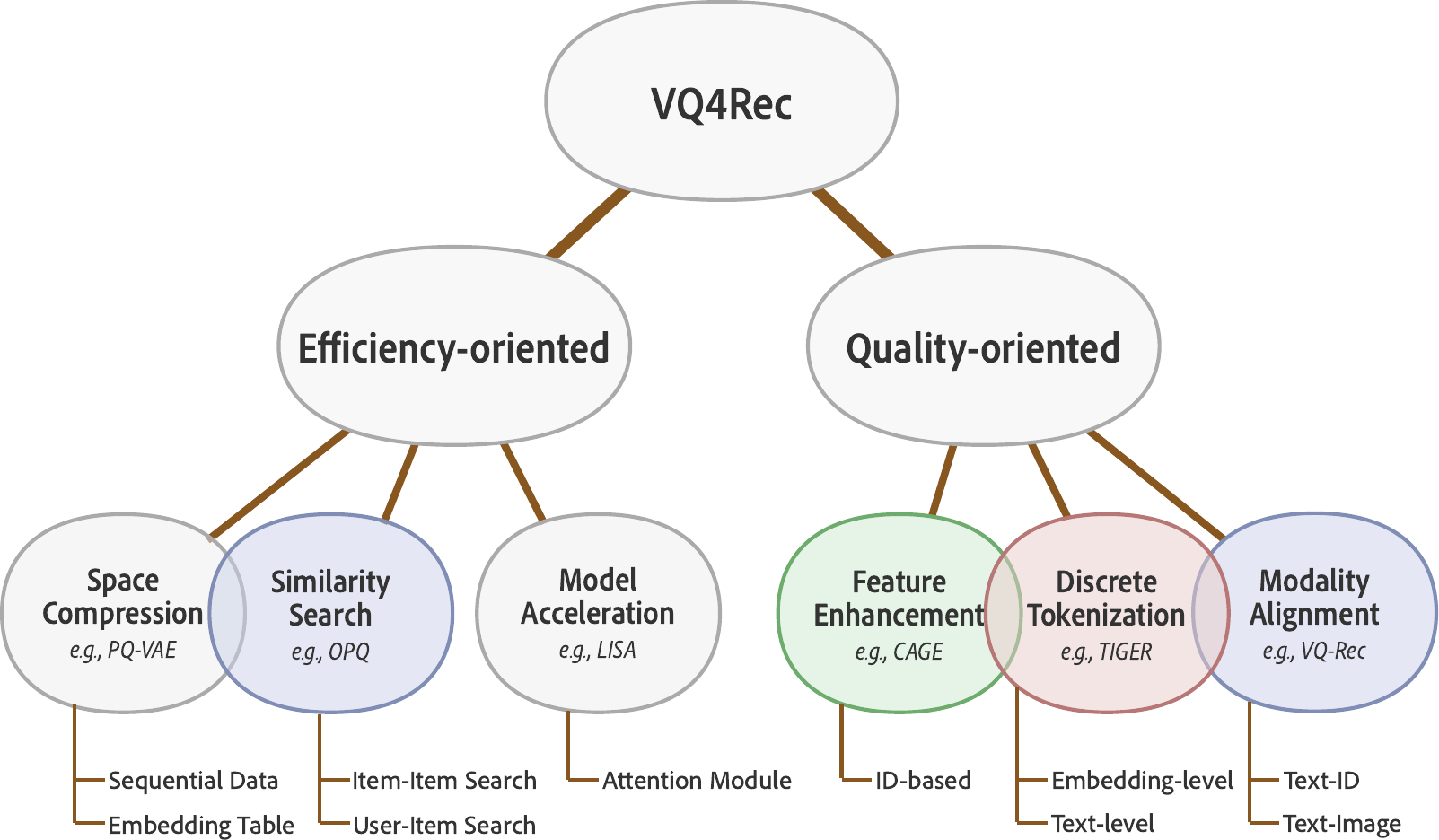}
    \caption{
    Categorization of VQ4Rec methods based on application scenario. The node colors denote different VQ techniques employed. The \tvq{standard}, \tpq{parallel}, and \tsq{sequential} VQ techniques are denoted by \textcolor{cvq}{green}, \textcolor{cpq}{blue}, and \textcolor{csq}{red}, respectively. The overlap between nodes indicates that the application scenarios they represent share certain similarities.
    }
    \label{fig:applications} 
\end{figure}

%% file: Content/Quality.tex
\section{Quality-oriented Approaches}

Building high-quality recommender systems is imperative to effectively cater to users' increasing information demands. 
Both academia and industry have explored various strategies to this end. These strategies include data augmentation, as demonstrated by \cite{song2022data}, which entails generating synthetic data from existing datasets through techniques like item masking \cite{slokom2019data}. Additionally, hyperparameter tuning, exemplified by \cite{wu2023hyperparameter}, automates the optimization of model settings, thereby mitigating the laborious process of grid search. Moreover, feature engineering, as elucidated by \cite{schifferer2020gpu}, enhances feature selection and data preprocessing.

VQ enhances recommender system quality by serving as a foundational step, specifically in item indexing for generative retrieval, a process which is further detailed in discrete tokenization applications. Furthermore, VQ aligns diverse modalities with soft constraints, facilitating multimodal feature learning.

\subsection{Feature Enhancement} 

Presently, recommender systems face challenges in cold-start scenarios where user interactions are sparse.
By integrating features such as item combination patterns and category information through VQ, these systems can be significantly enhanced.

To effectively utilize the data of active users, \citet{pan2021click} apply VQ to user interest clusters, facilitating cluster-level contrastive learning, which balances the personalization of representations between inactive and active users.
Their auto-quantized approach captures cluster-level similarities through VQ, in contrast to SimCLR proposed by ~\citet{chen2020simple}, which focuses solely on instance-level similarities.
To harness item combination patterns, \citet{luo2024within} propose VQA, which combines neural attention mechanism and VQ to determine the attention of candidate combination patterns.
To continuously generate and optimize the entity category trees over time, another study, CAGE~\citep{liu2024cage} enables the simultaneous learning and refinement of categorical code representations and entity embeddings in an end-to-end manner for ID-based recommendation.

However, these efforts rely on ID-based approaches, which may not be optimal in current diverse multimodal content landscape. Exploring methods to effectively leverage VQ techniques to enhance information from text, images, and other multimodal sources, and integrating it with recommendation features, presents a promising avenue for research.

\subsection{Modality Alignment}

Another interesting branch of work aims to improve modality alignment in recommender systems.
Transferable recommender systems are becoming increasingly important which can quickly adapt to new domains or scenarios. 
However, ensuring the alignment of various modalities and preserving their distinct patterns throughout downstream training models remains a challenge.

Under transferable scenario, VQ can be used for loosening the binding between item \emph{text} and \emph{ID }representation, as a sparse representation technique. \citet{hou2023learning} introduce VQ to represent items in a compact form, capturing the diverse characteristics of the products and addressing the transferability issues in sequential recommender systems. 
In contrast, \citet{hu2024lightweight} employ product quantization to impose additional modality constraints, targeting the mitigation of the modality forgetting issue in two-stage sequential recommenders. This involves transforming dissected \emph{text} and \emph{visual} correlations into discrete codebook representations to enforce tighter constraints.

Hence, VQ serves as a potent semantic bridge, particularly with the rise of Large Language Models (LLMs), facilitating connectivity across diverse modalities or domains. However, existing approaches primarily focus on aligning two modalities. Addressing multimodal scenarios involving more than three modalities necessitates novel solutions.

\subsection{Discrete Tokenization}

Tokenizing items and users in recommender systems has involved numerous strategies. Traditional methods often use atomic item identifiers (IDs), which can result in cold start problems. Later developments, inspired by document retrieval techniques like DSI~\cite{dsi} and NCI~\cite{nci}, introduced tree IDs using multi-layer K-Means~\cite{kmeans} to achieve discrete yet partially shared item tokens, though semantic discrepancies remained an issue. 

To address this, one line of research applies \emph{embedding-level reconstruction} task. For example, \citet{rajput2023recommender} developed the TIGER model based on RQ-VAE~\cite{rq-vae}, consisting of three steps: extracting item embeddings from content, discretizing these embeddings via residual quantization, and applying the discretized item tokens for sequence recommendation. Due to the inherent nature of residual quantization that can organise the tokens in a hierarchical manner, such approach proved highly successful and foundational for future research~\cite{singh2023better,liu2024semantic,liu2024mmgrec,jin2024contrastive}. Subsequent projects like LC-REC~\cite{zheng2023adapting} expanded on this by integrating item tokens into large models, hinting at the development of foundational recommendation models. Instead, some researchers~\cite{jin2023language} optimize this process further at \emph{text-level reconstruction} by treating item tokenization as a translation task within an encoder-decoder-decoder framework, using standard VQ on the top of the first decoder outputs that also achieves substantial performance.

However, the exploration of multimodal and multi-domain item tokenization remains limited, and this area presents a promising opportunity for advancing foundational recommender systems.

%% file: Content/Challenge.tex
\section{Future Directions}\label{sec:challenge}

In this section, we discuss the current challenges and emerging opportunities for future research in VQ4Rec.

\subsection{Codebook Collapse Problem}
There are some limitations associated with the capability of VQ.
For example, the challenge of codebook collapse may arise when only a minor portion of the codebook is effectively utilized.
VQ-VAE~\cite{vq-vae} employs STE~\cite{ste} to grant differentiability to VQ, consequently, many entries in the codebook remain unused or underutilized, restricting the model capacity to accurately represent and reconstruct input data. 
This core issue extends its impact to subsequent developments in recommender systems employing PQ-VAE~\cite{van2019pq} and RQ-VAE~\cite{rq-vae}, which impairs the recommender system's ability to offer varied and personalized recommendations to users as it fails to capture the diversity of the data.
At present, preliminary endeavors~\cite{zhang2023regularized,huh2023straightening,baykal2023edvae} have yielded encouraging results, with the scholarly community being urged to continue their research efforts in this direction.


\subsection{Item Discovery}
In item tokenization scenarios, the codebook space significantly exceeds the number of items in the dataset, suggesting that many potential code combinations remain untapped~\cite{rajput2023recommender}. Providing human-readable description for these new code combinations, especially in generative recommendation, represents a valuable direction. For instance, in product recommendations, this can help merchants develop products tailored to user demands; in video recommendations, it allows platforms to create personalized content based on the description. Currently, code training mainly relies on item embedding reconstruction tasks~\cite{zheng2023adapting,rajput2023recommender}. A viable alternative is an end-to-end reconstruction task based on item content such as title and description, where new code combinations are inputted into the decoder to generate the corresponding item content.

\subsection{User Tokenization}

Current VQ encoding schemes primarily focus on item discretization and have shown success in generative recommendation scenarios. However, discretizing user representation, i.e., user tokenization, also presents significant opportunities for research. For instance, \citet{liu2024semantic} has achieved substantial storage efficiency by applying discretization to both user and item in click through rate prediction. A pressing challenge is to enhance the quality of user tokens, which could enable large models to offer personalized responses through model personalization~\cite{ning2024user}.

\subsection{Multimodal Generative Recommendation}

Item semantic tokenization is currently the leading method for indexing items in generative recommender systems~\cite{rajput2023recommender,singh2023better,jin2023language}. However, current methods are mostly text-based, although multimodal semantic tokenization has begun to emerge in tasks such as text-to-image~\cite{zheng2024unicode} and video segmentation~\cite{xia2024achieving}. In the big data era, leveraging multimodal features offers a more comprehensive representation of items. Therefore, the development and application of multimodal tokenization techniques in recommender systems represents a critical advancement.

\subsection{RS--LLM Alignment}

The significant success of large language models~\cite{gpt-4} has established them as foundational elements across multiple fields. Current efforts increasingly focus on aligning object features from diverse domains with LLMs, enhancing their explainability and multimodal understanding~\cite{ge2023planting,zhan2024anygpt}. For example, LC-Rec~\cite{zheng2023adapting} has successfully finetuned discretized item IDs obtained by RQ-VAE~\cite{rq-vae} on the LLaMA model~\cite{llama-2}, validating this strategy in the recommendation domain. Future endeavors could involve integrating data from various domains to develop a foundational recommendation model with versatile skills.

\subsection{Codebook Quality Evaluation}


In some scenarios, the process of codebook generation and the recommendation task are not executed through end-to-end training. For instance, in item tokenization, item tokens are initially derived from item semantics before being evaluated in applications like sequential recommendation. Evaluating code quality through downstream tasks is both time-consuming and resource-intensive, suggesting a need for optimization. Therefore,  the exploration of methodologies for assessing code quality through the comparison of generated tokens against original inputs represents a significant and promising research direction.

\subsection{Efficient Large-scale Recommender Systems}

As large-scale models proliferate, the demand for efficient model training and inference is escalating within the recommendation community. 
VQ is emerging as a promising tool for enhancing the efficiency of large recommender systems, alongside other popular techniques like distillation and quantization. For instance, \citet{lingle2023transformer} and \citet{wu2021linear} have demonstrated that optimizing the attention mechanism through VQ can achieve linear time complexity in image generation and recommendation task, respectively.
However, these approaches typically involve smaller models and embedding dimensions that can be efficiently handled using a single codebook. In contrast, for larger models like LLaMA~\cite{llama-2}, which has embedding dimensions as large as 4096, the straightforward use of VQ may not be as effective. 
Exploring the integration of parallel quantization techniques with linear attention could potentially offer a viable solution.

%% file: Content/Conclusion.tex
\section{Conclusion}
VQ has become a pivotal element in the development of innovative solutions across various scenarios in recommender systems. With the advent of large language models, there has been a notable shift towards generative recommendation methods, where residual quantization has been widely adopted for its inherent advantages. 
However, the research of VQ4Rec is still in its early stage. This paper offers a comprehensive overview of current research in VQ4Rec, highlighting both efficiency-oriented and quality-oriented approaches. Additionally, we identify and discuss the open challenges and potential avenues for advancement. We hope this survey will foster continued exploration and innovation in VQ4Rec.



%% file: Content/Acknowledgement.tex
\section*{Acknowledgement}

Qijiong Liu is grateful to Prof. Min-Yen Kan from the National University of Singapore for his valuable comments and advice on this work during Liu's visit to NUS.

%% file: VQ4Rec.bbl

\begin{thebibliography}{94}


\ifx \showCODEN    \undefined \def \showCODEN     #1{\unskip}     \fi
\ifx \showDOI      \undefined \def \showDOI       #1{#1}\fi
\ifx \showISBNx    \undefined \def \showISBNx     #1{\unskip}     \fi
\ifx \showISBNxiii \undefined \def \showISBNxiii  #1{\unskip}     \fi
\ifx \showISSN     \undefined \def \showISSN      #1{\unskip}     \fi
\ifx \showLCCN     \undefined \def \showLCCN      #1{\unskip}     \fi
\ifx \shownote     \undefined \def \shownote      #1{#1}          \fi
\ifx \showarticletitle \undefined \def \showarticletitle #1{#1}   \fi
\ifx \showURL      \undefined \def \showURL       {\relax}        \fi
\providecommand\bibfield[2]{#2}
\providecommand\bibinfo[2]{#2}
\providecommand\natexlab[1]{#1}
\providecommand\showeprint[2][]{arXiv:#2}

\bibitem[Abe et~al\mbox{.}(1990)]%
        {abe1990voice}
\bibfield{author}{\bibinfo{person}{Masanobu Abe}, \bibinfo{person}{Satoshi Nakamura}, \bibinfo{person}{Kiyohiro Shikano}, {and} \bibinfo{person}{Hisao Kuwabara}.} \bibinfo{year}{1990}\natexlab{}.
\newblock \showarticletitle{Voice conversion through vector quantization}.
\newblock \bibinfo{journal}{\emph{Journal of the Acoustical Society of Japan (E)}} \bibinfo{volume}{11}, \bibinfo{number}{2} (\bibinfo{year}{1990}), \bibinfo{pages}{71--76}.
\newblock


\bibitem[Babenko and Lempitsky(2014)]%
        {aq}
\bibfield{author}{\bibinfo{person}{Artem Babenko} {and} \bibinfo{person}{Victor Lempitsky}.} \bibinfo{year}{2014}\natexlab{}.
\newblock \showarticletitle{Additive quantization for extreme vector compression}. In \bibinfo{booktitle}{\emph{Proceedings of the IEEE Conference on Computer Vision and Pattern Recognition}}. \bibinfo{pages}{931--938}.
\newblock


\bibitem[Baykal et~al\mbox{.}(2023)]%
        {baykal2023edvae}
\bibfield{author}{\bibinfo{person}{Gulcin Baykal}, \bibinfo{person}{Melih Kandemir}, {and} \bibinfo{person}{Gozde Unal}.} \bibinfo{year}{2023}\natexlab{}.
\newblock \showarticletitle{EdVAE: Mitigating Codebook Collapse with Evidential Discrete Variational Autoencoders}.
\newblock \bibinfo{journal}{\emph{Available at SSRN 4671725}} (\bibinfo{year}{2023}).
\newblock


\bibitem[Beel and Brunel(2019)]%
        {beel2019data}
\bibfield{author}{\bibinfo{person}{Joeran Beel} {and} \bibinfo{person}{Victor Brunel}.} \bibinfo{year}{2019}\natexlab{}.
\newblock \showarticletitle{Data pruning in recommender systems research: Best-practice or malpractice}.
\newblock \bibinfo{journal}{\emph{ACM RecSys}} (\bibinfo{year}{2019}).
\newblock


\bibitem[Bengio et~al\mbox{.}(2013)]%
        {ste}
\bibfield{author}{\bibinfo{person}{Yoshua Bengio}, \bibinfo{person}{Nicholas L{\'e}onard}, {and} \bibinfo{person}{Aaron Courville}.} \bibinfo{year}{2013}\natexlab{}.
\newblock \showarticletitle{Estimating or propagating gradients through stochastic neurons for conditional computation}.
\newblock \bibinfo{journal}{\emph{arXiv preprint arXiv:1308.3432}} (\bibinfo{year}{2013}).
\newblock


\bibitem[Buzo et~al\mbox{.}(1980)]%
        {buzo1980speech}
\bibfield{author}{\bibinfo{person}{Andr{\'e}s Buzo}, \bibinfo{person}{A Gray}, \bibinfo{person}{R Gray}, {and} \bibinfo{person}{John Markel}.} \bibinfo{year}{1980}\natexlab{}.
\newblock \showarticletitle{Speech coding based upon vector quantization}.
\newblock \bibinfo{journal}{\emph{IEEE Transactions on Acoustics, Speech, and Signal Processing}} \bibinfo{volume}{28}, \bibinfo{number}{5} (\bibinfo{year}{1980}), \bibinfo{pages}{562--574}.
\newblock


\bibitem[Cao et~al\mbox{.}(2017)]%
        {cao2017deep}
\bibfield{author}{\bibinfo{person}{Yue Cao}, \bibinfo{person}{Mingsheng Long}, \bibinfo{person}{Jianmin Wang}, {and} \bibinfo{person}{Shichen Liu}.} \bibinfo{year}{2017}\natexlab{}.
\newblock \showarticletitle{Deep visual-semantic quantization for efficient image retrieval}. In \bibinfo{booktitle}{\emph{Proceedings of the IEEE Conference on Computer Vision and Pattern Recognition}}. \bibinfo{pages}{1328--1337}.
\newblock


\bibitem[Chen et~al\mbox{.}(2022)]%
        {chen2022adversarial}
\bibfield{author}{\bibinfo{person}{Bin Chen}, \bibinfo{person}{Yan Feng}, \bibinfo{person}{Tao Dai}, \bibinfo{person}{Jiawang Bai}, \bibinfo{person}{Yong Jiang}, \bibinfo{person}{Shu-Tao Xia}, {and} \bibinfo{person}{Xuan Wang}.} \bibinfo{year}{2022}\natexlab{}.
\newblock \showarticletitle{Adversarial examples generation for deep product quantization networks on image retrieval}.
\newblock \bibinfo{journal}{\emph{IEEE Transactions on Pattern Analysis and Machine Intelligence}} \bibinfo{volume}{45}, \bibinfo{number}{2} (\bibinfo{year}{2022}), \bibinfo{pages}{1388--1404}.
\newblock


\bibitem[Chen et~al\mbox{.}(2023b)]%
        {chen2023hessian}
\bibfield{author}{\bibinfo{person}{Huiyuan Chen}, \bibinfo{person}{Kaixiong Zhou}, \bibinfo{person}{Kwei~Herng Lai}, \bibinfo{person}{Chin-Chia~Michael Yeh}, \bibinfo{person}{Yan Zheng}, \bibinfo{person}{Xia Hu}, {and} \bibinfo{person}{Hao Yang}.} \bibinfo{year}{2023}\natexlab{b}.
\newblock \showarticletitle{Hessian-aware Quantized Node Embeddings for Recommendation}. In \bibinfo{booktitle}{\emph{Proceedings of the 17th ACM Conference on Recommender Systems}}. \bibinfo{pages}{757--762}.
\newblock


\bibitem[Chen et~al\mbox{.}(2020)]%
        {chen2020simple}
\bibfield{author}{\bibinfo{person}{Ting Chen}, \bibinfo{person}{Simon Kornblith}, \bibinfo{person}{Mohammad Norouzi}, {and} \bibinfo{person}{Geoffrey Hinton}.} \bibinfo{year}{2020}\natexlab{}.
\newblock \showarticletitle{A simple framework for contrastive learning of visual representations}. In \bibinfo{booktitle}{\emph{International conference on machine learning}}. PMLR, \bibinfo{pages}{1597--1607}.
\newblock


\bibitem[Chen et~al\mbox{.}(2010)]%
        {chen2010approximate}
\bibfield{author}{\bibinfo{person}{Yongjian Chen}, \bibinfo{person}{Tao Guan}, {and} \bibinfo{person}{Cheng Wang}.} \bibinfo{year}{2010}\natexlab{}.
\newblock \showarticletitle{Approximate nearest neighbor search by residual vector quantization}.
\newblock \bibinfo{journal}{\emph{Sensors}} \bibinfo{volume}{10}, \bibinfo{number}{12} (\bibinfo{year}{2010}), \bibinfo{pages}{11259--11273}.
\newblock


\bibitem[Chen et~al\mbox{.}(2023a)]%
        {chen2023clustered}
\bibfield{author}{\bibinfo{person}{Yizhou Chen}, \bibinfo{person}{Guangda Huzhang}, \bibinfo{person}{Anxiang Zeng}, \bibinfo{person}{Qingtao Yu}, \bibinfo{person}{Hui Sun}, \bibinfo{person}{Hengyi Li}, \bibinfo{person}{Jingyi Li}, \bibinfo{person}{Yabo Ni}, \bibinfo{person}{Han Yu}, {and} \bibinfo{person}{Zhiming Zhou}.} \bibinfo{year}{2023}\natexlab{a}.
\newblock \showarticletitle{Clustered Embedding Learning for Recommender Systems}.
\newblock \bibinfo{journal}{\emph{arXiv preprint arXiv:2302.01478}} (\bibinfo{year}{2023}).
\newblock


\bibitem[Choromanski et~al\mbox{.}(2020)]%
        {performer}
\bibfield{author}{\bibinfo{person}{Krzysztof Choromanski}, \bibinfo{person}{Valerii Likhosherstov}, \bibinfo{person}{David Dohan}, \bibinfo{person}{Xingyou Song}, \bibinfo{person}{Andreea Gane}, \bibinfo{person}{Tamas Sarlos}, \bibinfo{person}{Peter Hawkins}, \bibinfo{person}{Jared Davis}, \bibinfo{person}{Afroz Mohiuddin}, \bibinfo{person}{Lukasz Kaiser}, {et~al\mbox{.}}} \bibinfo{year}{2020}\natexlab{}.
\newblock \showarticletitle{Rethinking attention with performers}.
\newblock \bibinfo{journal}{\emph{arXiv preprint arXiv:2009.14794}} (\bibinfo{year}{2020}).
\newblock


\bibitem[Cosman et~al\mbox{.}(1993)]%
        {cosman1993using}
\bibfield{author}{\bibinfo{person}{Pamela~C Cosman}, \bibinfo{person}{Karen~L Oehler}, \bibinfo{person}{Eve~A Riskin}, {and} \bibinfo{person}{Robert~M Gray}.} \bibinfo{year}{1993}\natexlab{}.
\newblock \showarticletitle{Using vector quantization for image processing}.
\newblock \bibinfo{journal}{\emph{Proc. IEEE}} \bibinfo{volume}{81}, \bibinfo{number}{9} (\bibinfo{year}{1993}), \bibinfo{pages}{1326--1341}.
\newblock


\bibitem[Dieleman et~al\mbox{.}(2021)]%
        {dieleman2021variable}
\bibfield{author}{\bibinfo{person}{Sander Dieleman}, \bibinfo{person}{Charlie Nash}, \bibinfo{person}{Jesse Engel}, {and} \bibinfo{person}{Karen Simonyan}.} \bibinfo{year}{2021}\natexlab{}.
\newblock \showarticletitle{Variable-rate discrete representation learning}.
\newblock \bibinfo{journal}{\emph{arXiv preprint arXiv:2103.06089}} (\bibinfo{year}{2021}).
\newblock


\bibitem[Donahue et~al\mbox{.}(2019)]%
        {donahue2019piano}
\bibfield{author}{\bibinfo{person}{Chris Donahue}, \bibinfo{person}{Ian Simon}, {and} \bibinfo{person}{Sander Dieleman}.} \bibinfo{year}{2019}\natexlab{}.
\newblock \showarticletitle{Piano genie}. In \bibinfo{booktitle}{\emph{Proceedings of the 24th International Conference on Intelligent User Interfaces}}. \bibinfo{pages}{160--164}.
\newblock


\bibitem[Feng et~al\mbox{.}(2023)]%
        {feng2023reinforcement}
\bibfield{author}{\bibinfo{person}{Chao Feng}, \bibinfo{person}{Defu Lian}, \bibinfo{person}{Xiting Wang}, \bibinfo{person}{Zheng Liu}, \bibinfo{person}{Xing Xie}, {and} \bibinfo{person}{Enhong Chen}.} \bibinfo{year}{2023}\natexlab{}.
\newblock \showarticletitle{Reinforcement routing on proximity graph for efficient recommendation}.
\newblock \bibinfo{journal}{\emph{ACM Transactions on Information Systems}} \bibinfo{volume}{41}, \bibinfo{number}{1} (\bibinfo{year}{2023}), \bibinfo{pages}{1--27}.
\newblock


\bibitem[Ge et~al\mbox{.}(2013)]%
        {opq}
\bibfield{author}{\bibinfo{person}{Tiezheng Ge}, \bibinfo{person}{Kaiming He}, \bibinfo{person}{Qifa Ke}, {and} \bibinfo{person}{Jian Sun}.} \bibinfo{year}{2013}\natexlab{}.
\newblock \showarticletitle{Optimized product quantization}.
\newblock \bibinfo{journal}{\emph{IEEE transactions on pattern analysis and machine intelligence}} \bibinfo{volume}{36}, \bibinfo{number}{4} (\bibinfo{year}{2013}), \bibinfo{pages}{744--755}.
\newblock


\bibitem[Ge et~al\mbox{.}(2023)]%
        {ge2023planting}
\bibfield{author}{\bibinfo{person}{Yuying Ge}, \bibinfo{person}{Yixiao Ge}, \bibinfo{person}{Ziyun Zeng}, \bibinfo{person}{Xintao Wang}, {and} \bibinfo{person}{Ying Shan}.} \bibinfo{year}{2023}\natexlab{}.
\newblock \showarticletitle{Planting a seed of vision in large language model}.
\newblock \bibinfo{journal}{\emph{arXiv preprint arXiv:2307.08041}} (\bibinfo{year}{2023}).
\newblock


\bibitem[Gray(1984)]%
        {vq}
\bibfield{author}{\bibinfo{person}{Robert Gray}.} \bibinfo{year}{1984}\natexlab{}.
\newblock \showarticletitle{Vector quantization}.
\newblock \bibinfo{journal}{\emph{IEEE Assp Magazine}} \bibinfo{volume}{1}, \bibinfo{number}{2} (\bibinfo{year}{1984}), \bibinfo{pages}{4--29}.
\newblock


\bibitem[Gray and Neuhoff(1998)]%
        {gray1998quantization}
\bibfield{author}{\bibinfo{person}{Robert~M. Gray} {and} \bibinfo{person}{David~L. Neuhoff}.} \bibinfo{year}{1998}\natexlab{}.
\newblock \showarticletitle{Quantization}.
\newblock \bibinfo{journal}{\emph{IEEE transactions on information theory}} \bibinfo{volume}{44}, \bibinfo{number}{6} (\bibinfo{year}{1998}), \bibinfo{pages}{2325--2383}.
\newblock


\bibitem[Hou et~al\mbox{.}(2023)]%
        {hou2023learning}
\bibfield{author}{\bibinfo{person}{Yupeng Hou}, \bibinfo{person}{Zhankui He}, \bibinfo{person}{Julian McAuley}, {and} \bibinfo{person}{Wayne~Xin Zhao}.} \bibinfo{year}{2023}\natexlab{}.
\newblock \showarticletitle{Learning vector-quantized item representation for transferable sequential recommenders}. In \bibinfo{booktitle}{\emph{Proceedings of the ACM Web Conference 2023}}. \bibinfo{pages}{1162--1171}.
\newblock


\bibitem[Hu et~al\mbox{.}(2021)]%
        {hu2021lora}
\bibfield{author}{\bibinfo{person}{Edward~J Hu}, \bibinfo{person}{Yelong Shen}, \bibinfo{person}{Phillip Wallis}, \bibinfo{person}{Zeyuan Allen-Zhu}, \bibinfo{person}{Yuanzhi Li}, \bibinfo{person}{Shean Wang}, \bibinfo{person}{Lu Wang}, {and} \bibinfo{person}{Weizhu Chen}.} \bibinfo{year}{2021}\natexlab{}.
\newblock \showarticletitle{Lora: Low-rank adaptation of large language models}.
\newblock \bibinfo{journal}{\emph{arXiv preprint arXiv:2106.09685}} (\bibinfo{year}{2021}).
\newblock


\bibitem[Hu et~al\mbox{.}(2024)]%
        {hu2024lightweight}
\bibfield{author}{\bibinfo{person}{Hengchang Hu}, \bibinfo{person}{Qijiong Liu}, \bibinfo{person}{Chuang Li}, {and} \bibinfo{person}{Min-Yen Kan}.} \bibinfo{year}{2024}\natexlab{}.
\newblock \showarticletitle{Lightweight Modality Adaptation to Sequential Recommendation via Correlation Supervision}. In \bibinfo{booktitle}{\emph{European Conference on Information Retrieval}}. \bibinfo{publisher}{Springer International Publishing}, \bibinfo{address}{Glasgow, Scotland, UK}.
\newblock


\bibitem[Huang and Jenor(2004)]%
        {huang2004audio}
\bibfield{author}{\bibinfo{person}{Yao-Chang Huang} {and} \bibinfo{person}{Shyh-Kang Jenor}.} \bibinfo{year}{2004}\natexlab{}.
\newblock \showarticletitle{An audio recommendation system based on audio signature description scheme in mpeg-7 audio}. In \bibinfo{booktitle}{\emph{2004 IEEE International Conference on Multimedia and Expo (ICME)(IEEE Cat. No. 04TH8763)}}, Vol.~\bibinfo{volume}{1}. IEEE, \bibinfo{pages}{639--642}.
\newblock


\bibitem[Huh et~al\mbox{.}(2023)]%
        {huh2023straightening}
\bibfield{author}{\bibinfo{person}{Minyoung Huh}, \bibinfo{person}{Brian Cheung}, \bibinfo{person}{Pulkit Agrawal}, {and} \bibinfo{person}{Phillip Isola}.} \bibinfo{year}{2023}\natexlab{}.
\newblock \showarticletitle{Straightening out the straight-through estimator: Overcoming optimization challenges in vector quantized networks}. In \bibinfo{booktitle}{\emph{International Conference on Machine Learning}}. PMLR, \bibinfo{pages}{14096--14113}.
\newblock


\bibitem[Imran et~al\mbox{.}(2023)]%
        {imran2023refrs}
\bibfield{author}{\bibinfo{person}{Mubashir Imran}, \bibinfo{person}{Hongzhi Yin}, \bibinfo{person}{Tong Chen}, \bibinfo{person}{Quoc Viet~Hung Nguyen}, \bibinfo{person}{Alexander Zhou}, {and} \bibinfo{person}{Kai Zheng}.} \bibinfo{year}{2023}\natexlab{}.
\newblock \showarticletitle{ReFRS: Resource-efficient federated recommender system for dynamic and diversified user preferences}.
\newblock \bibinfo{journal}{\emph{ACM Transactions on Information Systems}} \bibinfo{volume}{41}, \bibinfo{number}{3} (\bibinfo{year}{2023}), \bibinfo{pages}{1--30}.
\newblock


\bibitem[Jang and Cho(2021)]%
        {jang2021self}
\bibfield{author}{\bibinfo{person}{Young~Kyun Jang} {and} \bibinfo{person}{Nam~Ik Cho}.} \bibinfo{year}{2021}\natexlab{}.
\newblock \showarticletitle{Self-supervised product quantization for deep unsupervised image retrieval}. In \bibinfo{booktitle}{\emph{Proceedings of the IEEE/CVF international conference on computer vision}}. \bibinfo{pages}{12085--12094}.
\newblock


\bibitem[Jegou et~al\mbox{.}(2010)]%
        {pq}
\bibfield{author}{\bibinfo{person}{Herve Jegou}, \bibinfo{person}{Matthijs Douze}, {and} \bibinfo{person}{Cordelia Schmid}.} \bibinfo{year}{2010}\natexlab{}.
\newblock \showarticletitle{Product quantization for nearest neighbor search}.
\newblock \bibinfo{journal}{\emph{IEEE transactions on pattern analysis and machine intelligence}} \bibinfo{volume}{33}, \bibinfo{number}{1} (\bibinfo{year}{2010}), \bibinfo{pages}{117--128}.
\newblock


\bibitem[Jin et~al\mbox{.}(2023)]%
        {jin2023language}
\bibfield{author}{\bibinfo{person}{Bowen Jin}, \bibinfo{person}{Hansi Zeng}, \bibinfo{person}{Guoyin Wang}, \bibinfo{person}{Xiusi Chen}, \bibinfo{person}{Tianxin Wei}, \bibinfo{person}{Ruirui Li}, \bibinfo{person}{Zhengyang Wang}, \bibinfo{person}{Zheng Li}, \bibinfo{person}{Yang Li}, \bibinfo{person}{Hanqing Lu}, {et~al\mbox{.}}} \bibinfo{year}{2023}\natexlab{}.
\newblock \showarticletitle{Language Models As Semantic Indexers}.
\newblock \bibinfo{journal}{\emph{arXiv preprint arXiv:2310.07815}} (\bibinfo{year}{2023}).
\newblock


\bibitem[Jin et~al\mbox{.}(2024)]%
        {jin2024contrastive}
\bibfield{author}{\bibinfo{person}{Mengqun Jin}, \bibinfo{person}{Zexuan Qiu}, \bibinfo{person}{Jieming Zhu}, \bibinfo{person}{Zhenhua Dong}, {and} \bibinfo{person}{Xiu Li}.} \bibinfo{year}{2024}\natexlab{}.
\newblock \showarticletitle{Contrastive Quantization based Semantic Code for Generative Recommendation}.
\newblock \bibinfo{journal}{\emph{arXiv preprint arXiv:2404.14774}} (\bibinfo{year}{2024}).
\newblock


\bibitem[Johnson et~al\mbox{.}(2019)]%
        {johnson2019billion}
\bibfield{author}{\bibinfo{person}{Jeff Johnson}, \bibinfo{person}{Matthijs Douze}, {and} \bibinfo{person}{Herv{\'e} J{\'e}gou}.} \bibinfo{year}{2019}\natexlab{}.
\newblock \showarticletitle{Billion-scale similarity search with gpus}.
\newblock \bibinfo{journal}{\emph{IEEE Transactions on Big Data}} \bibinfo{volume}{7}, \bibinfo{number}{3} (\bibinfo{year}{2019}), \bibinfo{pages}{535--547}.
\newblock


\bibitem[Juang and Gray(1982)]%
        {juang1982multiple}
\bibfield{author}{\bibinfo{person}{Biing-Hwang Juang} {and} \bibinfo{person}{A Gray}.} \bibinfo{year}{1982}\natexlab{}.
\newblock \showarticletitle{Multiple stage vector quantization for speech coding}. In \bibinfo{booktitle}{\emph{ICASSP'82. IEEE International Conference on Acoustics, Speech, and Signal Processing}}, Vol.~\bibinfo{volume}{7}. IEEE, \bibinfo{pages}{597--600}.
\newblock


\bibitem[Kang et~al\mbox{.}(2020)]%
        {kang2020learning}
\bibfield{author}{\bibinfo{person}{Wang-Cheng Kang}, \bibinfo{person}{Derek~Zhiyuan Cheng}, \bibinfo{person}{Ting Chen}, \bibinfo{person}{Xinyang Yi}, \bibinfo{person}{Dong Lin}, \bibinfo{person}{Lichan Hong}, {and} \bibinfo{person}{Ed~H Chi}.} \bibinfo{year}{2020}\natexlab{}.
\newblock \showarticletitle{Learning multi-granular quantized embeddings for large-vocab categorical features in recommender systems}. In \bibinfo{booktitle}{\emph{Companion Proceedings of the Web Conference 2020}}. \bibinfo{pages}{562--566}.
\newblock


\bibitem[Ko et~al\mbox{.}(2021)]%
        {ko2021mascot}
\bibfield{author}{\bibinfo{person}{Yunyong Ko}, \bibinfo{person}{Jae-Seo Yu}, \bibinfo{person}{Hong-Kyun Bae}, \bibinfo{person}{Yongjun Park}, \bibinfo{person}{Dongwon Lee}, {and} \bibinfo{person}{Sang-Wook Kim}.} \bibinfo{year}{2021}\natexlab{}.
\newblock \showarticletitle{MASCOT: A Quantization Framework for Efficient Matrix Factorization in Recommender Systems}. In \bibinfo{booktitle}{\emph{2021 IEEE International Conference on Data Mining (ICDM)}}. IEEE, \bibinfo{pages}{290--299}.
\newblock


\bibitem[Koren et~al\mbox{.}(2009)]%
        {koren2009matrix}
\bibfield{author}{\bibinfo{person}{Yehuda Koren}, \bibinfo{person}{Robert Bell}, {and} \bibinfo{person}{Chris Volinsky}.} \bibinfo{year}{2009}\natexlab{}.
\newblock \showarticletitle{Matrix factorization techniques for recommender systems}.
\newblock \bibinfo{journal}{\emph{Computer}} \bibinfo{volume}{42}, \bibinfo{number}{8} (\bibinfo{year}{2009}), \bibinfo{pages}{30--37}.
\newblock


\bibitem[Krishna and Murty(1999)]%
        {kmeans}
\bibfield{author}{\bibinfo{person}{K Krishna} {and} \bibinfo{person}{M~Narasimha Murty}.} \bibinfo{year}{1999}\natexlab{}.
\newblock \showarticletitle{Genetic K-means algorithm}.
\newblock \bibinfo{journal}{\emph{IEEE Transactions on Systems, Man, and Cybernetics, Part B (Cybernetics)}} \bibinfo{volume}{29}, \bibinfo{number}{3} (\bibinfo{year}{1999}), \bibinfo{pages}{433--439}.
\newblock


\bibitem[{\L}a{\'n}cucki et~al\mbox{.}(2020)]%
        {ema}
\bibfield{author}{\bibinfo{person}{Adrian {\L}a{\'n}cucki}, \bibinfo{person}{Jan Chorowski}, \bibinfo{person}{Guillaume Sanchez}, \bibinfo{person}{Ricard Marxer}, \bibinfo{person}{Nanxin Chen}, \bibinfo{person}{Hans~JGA Dolfing}, \bibinfo{person}{Sameer Khurana}, \bibinfo{person}{Tanel Alum{\"a}e}, {and} \bibinfo{person}{Antoine Laurent}.} \bibinfo{year}{2020}\natexlab{}.
\newblock \showarticletitle{Robust training of vector quantized bottleneck models}. In \bibinfo{booktitle}{\emph{2020 International Joint Conference on Neural Networks (IJCNN)}}. IEEE, \bibinfo{pages}{1--7}.
\newblock


\bibitem[Lee et~al\mbox{.}(2022)]%
        {rq-vae}
\bibfield{author}{\bibinfo{person}{Doyup Lee}, \bibinfo{person}{Chiheon Kim}, \bibinfo{person}{Saehoon Kim}, \bibinfo{person}{Minsu Cho}, {and} \bibinfo{person}{Wook-Shin Han}.} \bibinfo{year}{2022}\natexlab{}.
\newblock \showarticletitle{Autoregressive image generation using residual quantization}. In \bibinfo{booktitle}{\emph{Proceedings of the IEEE/CVF Conference on Computer Vision and Pattern Recognition}}. \bibinfo{pages}{11523--11532}.
\newblock


\bibitem[Lian et~al\mbox{.}(2020a)]%
        {lian2020lightrec}
\bibfield{author}{\bibinfo{person}{Defu Lian}, \bibinfo{person}{Haoyu Wang}, \bibinfo{person}{Zheng Liu}, \bibinfo{person}{Jianxun Lian}, \bibinfo{person}{Enhong Chen}, {and} \bibinfo{person}{Xing Xie}.} \bibinfo{year}{2020}\natexlab{a}.
\newblock \showarticletitle{Lightrec: A memory and search-efficient recommender system}. In \bibinfo{booktitle}{\emph{Proceedings of The Web Conference 2020}}. \bibinfo{pages}{695--705}.
\newblock


\bibitem[Lian et~al\mbox{.}(2020b)]%
        {lian2020product}
\bibfield{author}{\bibinfo{person}{Defu Lian}, \bibinfo{person}{Xing Xie}, \bibinfo{person}{Enhong Chen}, {and} \bibinfo{person}{Hui Xiong}.} \bibinfo{year}{2020}\natexlab{b}.
\newblock \showarticletitle{Product quantized collaborative filtering}.
\newblock \bibinfo{journal}{\emph{IEEE Transactions on Knowledge and Data Engineering}} \bibinfo{volume}{33}, \bibinfo{number}{9} (\bibinfo{year}{2020}), \bibinfo{pages}{3284--3296}.
\newblock


\bibitem[Lingle(2023)]%
        {lingle2023transformer}
\bibfield{author}{\bibinfo{person}{Lucas~D Lingle}.} \bibinfo{year}{2023}\natexlab{}.
\newblock \showarticletitle{Transformer-vq: Linear-time transformers via vector quantization}.
\newblock \bibinfo{journal}{\emph{arXiv preprint arXiv:2309.16354}} (\bibinfo{year}{2023}).
\newblock


\bibitem[Liu et~al\mbox{.}(2022b)]%
        {liu2022few}
\bibfield{author}{\bibinfo{person}{Haokun Liu}, \bibinfo{person}{Derek Tam}, \bibinfo{person}{Mohammed Muqeeth}, \bibinfo{person}{Jay Mohta}, \bibinfo{person}{Tenghao Huang}, \bibinfo{person}{Mohit Bansal}, {and} \bibinfo{person}{Colin~A Raffel}.} \bibinfo{year}{2022}\natexlab{b}.
\newblock \showarticletitle{Few-shot parameter-efficient fine-tuning is better and cheaper than in-context learning}.
\newblock \bibinfo{journal}{\emph{Advances in Neural Information Processing Systems}}  \bibinfo{volume}{35} (\bibinfo{year}{2022}), \bibinfo{pages}{1950--1965}.
\newblock


\bibitem[Liu et~al\mbox{.}(2024d)]%
        {liu2024mmgrec}
\bibfield{author}{\bibinfo{person}{Han Liu}, \bibinfo{person}{Yinwei Wei}, \bibinfo{person}{Xuemeng Song}, \bibinfo{person}{Weili Guan}, \bibinfo{person}{Yuan-Fang Li}, {and} \bibinfo{person}{Liqiang Nie}.} \bibinfo{year}{2024}\natexlab{d}.
\newblock \showarticletitle{MMGRec: Multimodal Generative Recommendation with Transformer Model}.
\newblock \bibinfo{journal}{\emph{arXiv preprint arXiv:2404.16555}} (\bibinfo{year}{2024}).
\newblock


\bibitem[Liu et~al\mbox{.}(2022a)]%
        {ecoformer}
\bibfield{author}{\bibinfo{person}{Jing Liu}, \bibinfo{person}{Zizheng Pan}, \bibinfo{person}{Haoyu He}, \bibinfo{person}{Jianfei Cai}, {and} \bibinfo{person}{Bohan Zhuang}.} \bibinfo{year}{2022}\natexlab{a}.
\newblock \showarticletitle{Ecoformer: Energy-saving attention with linear complexity}.
\newblock \bibinfo{journal}{\emph{Advances in Neural Information Processing Systems}}  \bibinfo{volume}{35} (\bibinfo{year}{2022}), \bibinfo{pages}{10295--10308}.
\newblock


\bibitem[Liu et~al\mbox{.}(2023)]%
        {liu2023linrec}
\bibfield{author}{\bibinfo{person}{Langming Liu}, \bibinfo{person}{Liu Cai}, \bibinfo{person}{Chi Zhang}, \bibinfo{person}{Xiangyu Zhao}, \bibinfo{person}{Jingtong Gao}, \bibinfo{person}{Wanyu Wang}, \bibinfo{person}{Yifu Lv}, \bibinfo{person}{Wenqi Fan}, \bibinfo{person}{Yiqi Wang}, \bibinfo{person}{Ming He}, {et~al\mbox{.}}} \bibinfo{year}{2023}\natexlab{}.
\newblock \showarticletitle{Linrec: Linear attention mechanism for long-term sequential recommender systems}. In \bibinfo{booktitle}{\emph{Proceedings of the 46th International ACM SIGIR Conference on Research and Development in Information Retrieval}}. \bibinfo{pages}{289--299}.
\newblock


\bibitem[Liu et~al\mbox{.}(2024a)]%
        {liu2024once}
\bibfield{author}{\bibinfo{person}{Qijiong Liu}, \bibinfo{person}{Nuo Chen}, \bibinfo{person}{Tetsuya Sakai}, {and} \bibinfo{person}{Xiao-Ming Wu}.} \bibinfo{year}{2024}\natexlab{a}.
\newblock \showarticletitle{Once: Boosting content-based recommendation with both open-and closed-source large language models}. In \bibinfo{booktitle}{\emph{Proceedings of the 17th ACM International Conference on Web Search and Data Mining}}. \bibinfo{pages}{452--461}.
\newblock


\bibitem[Liu et~al\mbox{.}(2024b)]%
        {liu2024cage}
\bibfield{author}{\bibinfo{person}{Qijiong Liu}, \bibinfo{person}{Lu Fan}, \bibinfo{person}{Jiaren Xiao}, \bibinfo{person}{Jieming Zhu}, {and} \bibinfo{person}{Xiao-Ming Wu}.} \bibinfo{year}{2024}\natexlab{b}.
\newblock \showarticletitle{Learning Category Trees for ID-Based Recommendation: Exploring the Power of Differentiable Vector Quantization}. In \bibinfo{booktitle}{\emph{Proceedings of the ACM Web Conference 2024}}. \bibinfo{address}{Singapore}.
\newblock


\bibitem[Liu et~al\mbox{.}(2024c)]%
        {liu2024semantic}
\bibfield{author}{\bibinfo{person}{Qijiong Liu}, \bibinfo{person}{Hengchang Hu}, \bibinfo{person}{Jiahao Wu}, \bibinfo{person}{Jieming Zhu}, \bibinfo{person}{Min-Yen Kan}, {and} \bibinfo{person}{Xiao-Ming Wu}.} \bibinfo{year}{2024}\natexlab{c}.
\newblock \bibinfo{title}{Discrete Semantic Tokenization for Deep CTR Prediction}.
\newblock
\newblock


\bibitem[Liu et~al\mbox{.}(2024e)]%
        {liu2024benchmarking}
\bibfield{author}{\bibinfo{person}{Qijiong Liu}, \bibinfo{person}{Jieming Zhu}, \bibinfo{person}{Quanyu Dai}, {and} \bibinfo{person}{Xiao-Ming Wu}.} \bibinfo{year}{2024}\natexlab{e}.
\newblock \showarticletitle{Benchmarking News Recommendation in the Era of Green AI}.
\newblock \bibinfo{journal}{\emph{arXiv preprint arXiv:2403.04736}} (\bibinfo{year}{2024}).
\newblock


\bibitem[Lu and Teng(1999)]%
        {lu1999novel}
\bibfield{author}{\bibinfo{person}{Guojun Lu} {and} \bibinfo{person}{Shyhwei Teng}.} \bibinfo{year}{1999}\natexlab{}.
\newblock \showarticletitle{A novel image retrieval technique based on vector quantization}. In \bibinfo{booktitle}{\emph{Proceedings of International Conference on Computational Intelligence for Modeling, Control and Automation}}. Citeseer, \bibinfo{pages}{36--41}.
\newblock


\bibitem[Lu et~al\mbox{.}(2023)]%
        {lu2023differentiable}
\bibfield{author}{\bibinfo{person}{Zepu Lu}, \bibinfo{person}{Defu Lian}, \bibinfo{person}{Jin Zhang}, \bibinfo{person}{Zaixi Zhang}, \bibinfo{person}{Chao Feng}, \bibinfo{person}{Hao Wang}, {and} \bibinfo{person}{Enhong Chen}.} \bibinfo{year}{2023}\natexlab{}.
\newblock \showarticletitle{Differentiable Optimized Product Quantization and Beyond}. In \bibinfo{booktitle}{\emph{Proceedings of the ACM Web Conference 2023}}. \bibinfo{pages}{3353--3363}.
\newblock


\bibitem[Luo et~al\mbox{.}(2024)]%
        {luo2024within}
\bibfield{author}{\bibinfo{person}{Kai Luo}, \bibinfo{person}{Tianshu Shen}, \bibinfo{person}{Lan Yao}, \bibinfo{person}{Ga Wu}, \bibinfo{person}{Aaron Liblong}, \bibinfo{person}{Istvan Fehervari}, \bibinfo{person}{Ruijian An}, \bibinfo{person}{Jawad Ahmed}, \bibinfo{person}{Harshit Mishra}, {and} \bibinfo{person}{Charu Pujari}.} \bibinfo{year}{2024}\natexlab{}.
\newblock \showarticletitle{Within-basket Recommendation via Neural Pattern Associator}.
\newblock \bibinfo{journal}{\emph{arXiv preprint arXiv:2401.16433}} (\bibinfo{year}{2024}).
\newblock


\bibitem[Makhoul et~al\mbox{.}(1985)]%
        {makhoul1985vector}
\bibfield{author}{\bibinfo{person}{John Makhoul}, \bibinfo{person}{Salim Roucos}, {and} \bibinfo{person}{Herbert Gish}.} \bibinfo{year}{1985}\natexlab{}.
\newblock \showarticletitle{Vector quantization in speech coding}.
\newblock \bibinfo{journal}{\emph{Proc. IEEE}} \bibinfo{volume}{73}, \bibinfo{number}{11} (\bibinfo{year}{1985}), \bibinfo{pages}{1551--1588}.
\newblock


\bibitem[Martinez et~al\mbox{.}(2014)]%
        {rq}
\bibfield{author}{\bibinfo{person}{Julieta Martinez}, \bibinfo{person}{Holger~H Hoos}, {and} \bibinfo{person}{James~J Little}.} \bibinfo{year}{2014}\natexlab{}.
\newblock \showarticletitle{Stacked quantizers for compositional vector compression}.
\newblock \bibinfo{journal}{\emph{arXiv preprint arXiv:1411.2173}} (\bibinfo{year}{2014}).
\newblock


\bibitem[Menghani(2023)]%
        {menghani2023efficient}
\bibfield{author}{\bibinfo{person}{Gaurav Menghani}.} \bibinfo{year}{2023}\natexlab{}.
\newblock \showarticletitle{Efficient deep learning: A survey on making deep learning models smaller, faster, and better}.
\newblock \bibinfo{journal}{\emph{Comput. Surveys}} \bibinfo{volume}{55}, \bibinfo{number}{12} (\bibinfo{year}{2023}), \bibinfo{pages}{1--37}.
\newblock


\bibitem[Mentzer et~al\mbox{.}(2023)]%
        {mentzer2023finite}
\bibfield{author}{\bibinfo{person}{Fabian Mentzer}, \bibinfo{person}{David Minnen}, \bibinfo{person}{Eirikur Agustsson}, {and} \bibinfo{person}{Michael Tschannen}.} \bibinfo{year}{2023}\natexlab{}.
\newblock \showarticletitle{Finite scalar quantization: Vq-vae made simple}.
\newblock \bibinfo{journal}{\emph{arXiv preprint arXiv:2309.15505}} (\bibinfo{year}{2023}).
\newblock


\bibitem[Morozov and Babenko(2018)]%
        {morozov2018non}
\bibfield{author}{\bibinfo{person}{Stanislav Morozov} {and} \bibinfo{person}{Artem Babenko}.} \bibinfo{year}{2018}\natexlab{}.
\newblock \showarticletitle{Non-metric similarity graphs for maximum inner product search}.
\newblock \bibinfo{journal}{\emph{Advances in Neural Information Processing Systems}}  \bibinfo{volume}{31} (\bibinfo{year}{2018}).
\newblock


\bibitem[Nasrabadi and King(1988)]%
        {nasrabadi1988image}
\bibfield{author}{\bibinfo{person}{Nasser~M Nasrabadi} {and} \bibinfo{person}{Robert~A King}.} \bibinfo{year}{1988}\natexlab{}.
\newblock \showarticletitle{Image coding using vector quantization: A review}.
\newblock \bibinfo{journal}{\emph{IEEE Transactions on communications}} \bibinfo{volume}{36}, \bibinfo{number}{8} (\bibinfo{year}{1988}), \bibinfo{pages}{957--971}.
\newblock


\bibitem[Neyshabur and Srebro(2015)]%
        {neyshabur2015symmetric}
\bibfield{author}{\bibinfo{person}{Behnam Neyshabur} {and} \bibinfo{person}{Nathan Srebro}.} \bibinfo{year}{2015}\natexlab{}.
\newblock \showarticletitle{On symmetric and asymmetric lshs for inner product search}. In \bibinfo{booktitle}{\emph{International Conference on Machine Learning}}. PMLR, \bibinfo{pages}{1926--1934}.
\newblock


\bibitem[Ning et~al\mbox{.}(2024)]%
        {ning2024user}
\bibfield{author}{\bibinfo{person}{Lin Ning}, \bibinfo{person}{Luyang Liu}, \bibinfo{person}{Jiaxing Wu}, \bibinfo{person}{Neo Wu}, \bibinfo{person}{Devora Berlowitz}, \bibinfo{person}{Sushant Prakash}, \bibinfo{person}{Bradley Green}, \bibinfo{person}{Shawn O'Banion}, {and} \bibinfo{person}{Jun Xie}.} \bibinfo{year}{2024}\natexlab{}.
\newblock \showarticletitle{User-LLM: Efficient LLM Contextualization with User Embeddings}.
\newblock \bibinfo{journal}{\emph{arXiv preprint arXiv:2402.13598}} (\bibinfo{year}{2024}).
\newblock


\bibitem[OpenAI(2023)]%
        {gpt-4}
\bibfield{author}{\bibinfo{person}{R OpenAI}.} \bibinfo{year}{2023}\natexlab{}.
\newblock \showarticletitle{Gpt-4 technical report. arxiv 2303.08774}.
\newblock \bibinfo{journal}{\emph{View in Article}} \bibinfo{volume}{2}, \bibinfo{number}{5} (\bibinfo{year}{2023}).
\newblock


\bibitem[Pan et~al\mbox{.}(2021)]%
        {pan2021click}
\bibfield{author}{\bibinfo{person}{Yujie Pan}, \bibinfo{person}{Jiangchao Yao}, \bibinfo{person}{Bo Han}, \bibinfo{person}{Kunyang Jia}, \bibinfo{person}{Ya Zhang}, {and} \bibinfo{person}{Hongxia Yang}.} \bibinfo{year}{2021}\natexlab{}.
\newblock \showarticletitle{Click-through rate prediction with auto-quantized contrastive learning}.
\newblock \bibinfo{journal}{\emph{arXiv preprint arXiv:2109.13921}} (\bibinfo{year}{2021}).
\newblock


\bibitem[Rajput et~al\mbox{.}(2023)]%
        {rajput2023recommender}
\bibfield{author}{\bibinfo{person}{Shashank Rajput}, \bibinfo{person}{Nikhil Mehta}, \bibinfo{person}{Anima Singh}, \bibinfo{person}{Raghunandan~H Keshavan}, \bibinfo{person}{Trung Vu}, \bibinfo{person}{Lukasz Heldt}, \bibinfo{person}{Lichan Hong}, \bibinfo{person}{Yi Tay}, \bibinfo{person}{Vinh~Q Tran}, \bibinfo{person}{Jonah Samost}, {et~al\mbox{.}}} \bibinfo{year}{2023}\natexlab{}.
\newblock \showarticletitle{Recommender Systems with Generative Retrieval}.
\newblock \bibinfo{journal}{\emph{arXiv preprint arXiv:2305.05065}} (\bibinfo{year}{2023}).
\newblock


\bibitem[Razavi et~al\mbox{.}(2019)]%
        {razavi2019generating}
\bibfield{author}{\bibinfo{person}{Ali Razavi}, \bibinfo{person}{Aaron Van~den Oord}, {and} \bibinfo{person}{Oriol Vinyals}.} \bibinfo{year}{2019}\natexlab{}.
\newblock \showarticletitle{Generating diverse high-fidelity images with vq-vae-2}.
\newblock \bibinfo{journal}{\emph{Advances in neural information processing systems}}  \bibinfo{volume}{32} (\bibinfo{year}{2019}).
\newblock


\bibitem[Sabin and Gray(1984)]%
        {sabin1984product}
\bibfield{author}{\bibinfo{person}{ML Sabin} {and} \bibinfo{person}{R Gray}.} \bibinfo{year}{1984}\natexlab{}.
\newblock \showarticletitle{Product code vector quantizers for waveform and voice coding}.
\newblock \bibinfo{journal}{\emph{IEEE transactions on acoustics, speech, and signal processing}} \bibinfo{volume}{32}, \bibinfo{number}{3} (\bibinfo{year}{1984}), \bibinfo{pages}{474--488}.
\newblock


\bibitem[Schifferer et~al\mbox{.}(2020)]%
        {schifferer2020gpu}
\bibfield{author}{\bibinfo{person}{Benedikt Schifferer}, \bibinfo{person}{Gilberto Titericz}, \bibinfo{person}{Chris Deotte}, \bibinfo{person}{Christof Henkel}, \bibinfo{person}{Kazuki Onodera}, \bibinfo{person}{Jiwei Liu}, \bibinfo{person}{Bojan Tunguz}, \bibinfo{person}{Even Oldridge}, \bibinfo{person}{Gabriel De~Souza Pereira~Moreira}, {and} \bibinfo{person}{Ahmet Erdem}.} \bibinfo{year}{2020}\natexlab{}.
\newblock \showarticletitle{GPU accelerated feature engineering and training for recommender systems}.
\newblock In \bibinfo{booktitle}{\emph{Proceedings of the Recommender Systems Challenge 2020}}. \bibinfo{pages}{16--23}.
\newblock


\bibitem[Schwartz et~al\mbox{.}(2020)]%
        {schwartz2020green}
\bibfield{author}{\bibinfo{person}{Roy Schwartz}, \bibinfo{person}{Jesse Dodge}, \bibinfo{person}{Noah~A Smith}, {and} \bibinfo{person}{Oren Etzioni}.} \bibinfo{year}{2020}\natexlab{}.
\newblock \showarticletitle{Green ai}.
\newblock \bibinfo{journal}{\emph{Commun. ACM}} \bibinfo{volume}{63}, \bibinfo{number}{12} (\bibinfo{year}{2020}), \bibinfo{pages}{54--63}.
\newblock


\bibitem[Shi et~al\mbox{.}(2023)]%
        {shi2023quantize}
\bibfield{author}{\bibinfo{person}{Lingfeng Shi}, \bibinfo{person}{Yuang Liu}, \bibinfo{person}{Jun Wang}, {and} \bibinfo{person}{Wei Zhang}.} \bibinfo{year}{2023}\natexlab{}.
\newblock \showarticletitle{Quantize Sequential Recommenders Without Private Data}. In \bibinfo{booktitle}{\emph{Proceedings of the ACM Web Conference 2023}}. \bibinfo{pages}{1043--1052}.
\newblock


\bibitem[Singh et~al\mbox{.}(2023)]%
        {singh2023better}
\bibfield{author}{\bibinfo{person}{Anima Singh}, \bibinfo{person}{Trung Vu}, \bibinfo{person}{Raghunandan Keshavan}, \bibinfo{person}{Nikhil Mehta}, \bibinfo{person}{Xinyang Yi}, \bibinfo{person}{Lichan Hong}, \bibinfo{person}{Lukasz Heldt}, \bibinfo{person}{Li Wei}, \bibinfo{person}{Ed Chi}, {and} \bibinfo{person}{Maheswaran Sathiamoorthy}.} \bibinfo{year}{2023}\natexlab{}.
\newblock \showarticletitle{Better Generalization with Semantic IDs: A case study in Ranking for Recommendations}.
\newblock \bibinfo{journal}{\emph{arXiv preprint arXiv:2306.08121}} (\bibinfo{year}{2023}).
\newblock


\bibitem[Slokom et~al\mbox{.}(2019)]%
        {slokom2019data}
\bibfield{author}{\bibinfo{person}{Manel Slokom}, \bibinfo{person}{Martha Larson}, {and} \bibinfo{person}{Alan Hanjalic}.} \bibinfo{year}{2019}\natexlab{}.
\newblock \showarticletitle{Data masking for recommender systems: prediction performance and rating hiding}.
\newblock  (\bibinfo{year}{2019}).
\newblock


\bibitem[Song and Suh(2022)]%
        {song2022data}
\bibfield{author}{\bibinfo{person}{Joo-yeong Song} {and} \bibinfo{person}{Bongwon Suh}.} \bibinfo{year}{2022}\natexlab{}.
\newblock \showarticletitle{Data Augmentation Strategies for Improving Sequential Recommender Systems}.
\newblock \bibinfo{journal}{\emph{arXiv e-prints}} (\bibinfo{year}{2022}), \bibinfo{pages}{arXiv--2203}.
\newblock


\bibitem[Su et~al\mbox{.}(2023)]%
        {su2023beyond}
\bibfield{author}{\bibinfo{person}{Liangcai Su}, \bibinfo{person}{Fan Yan}, \bibinfo{person}{Jieming Zhu}, \bibinfo{person}{Xi Xiao}, \bibinfo{person}{Haoyi Duan}, \bibinfo{person}{Zhou Zhao}, \bibinfo{person}{Zhenhua Dong}, {and} \bibinfo{person}{Ruiming Tang}.} \bibinfo{year}{2023}\natexlab{}.
\newblock \showarticletitle{Beyond Two-Tower Matching: Learning Sparse Retrievable Cross-Interactions for Recommendation}. In \bibinfo{booktitle}{\emph{Proceedings of the 46th International ACM SIGIR Conference on Research and Development in Information Retrieval}}. \bibinfo{pages}{548--557}.
\newblock


\bibitem[Tang and Wang(2018)]%
        {tang2018ranking}
\bibfield{author}{\bibinfo{person}{Jiaxi Tang} {and} \bibinfo{person}{Ke Wang}.} \bibinfo{year}{2018}\natexlab{}.
\newblock \showarticletitle{Ranking distillation: Learning compact ranking models with high performance for recommender system}. In \bibinfo{booktitle}{\emph{Proceedings of the 24th ACM SIGKDD international conference on knowledge discovery \& data mining}}. \bibinfo{pages}{2289--2298}.
\newblock


\bibitem[Tay et~al\mbox{.}(2022)]%
        {dsi}
\bibfield{author}{\bibinfo{person}{Yi Tay}, \bibinfo{person}{Vinh Tran}, \bibinfo{person}{Mostafa Dehghani}, \bibinfo{person}{Jianmo Ni}, \bibinfo{person}{Dara Bahri}, \bibinfo{person}{Harsh Mehta}, \bibinfo{person}{Zhen Qin}, \bibinfo{person}{Kai Hui}, \bibinfo{person}{Zhe Zhao}, \bibinfo{person}{Jai Gupta}, {et~al\mbox{.}}} \bibinfo{year}{2022}\natexlab{}.
\newblock \showarticletitle{Transformer memory as a differentiable search index}.
\newblock \bibinfo{journal}{\emph{Advances in Neural Information Processing Systems}}  \bibinfo{volume}{35} (\bibinfo{year}{2022}), \bibinfo{pages}{21831--21843}.
\newblock


\bibitem[Touvron et~al\mbox{.}(2023)]%
        {llama-2}
\bibfield{author}{\bibinfo{person}{Hugo Touvron}, \bibinfo{person}{Louis Martin}, \bibinfo{person}{Kevin Stone}, \bibinfo{person}{Peter Albert}, \bibinfo{person}{Amjad Almahairi}, \bibinfo{person}{Yasmine Babaei}, \bibinfo{person}{Nikolay Bashlykov}, \bibinfo{person}{Soumya Batra}, \bibinfo{person}{Prajjwal Bhargava}, \bibinfo{person}{Shruti Bhosale}, {et~al\mbox{.}}} \bibinfo{year}{2023}\natexlab{}.
\newblock \showarticletitle{Llama 2: Open foundation and fine-tuned chat models}.
\newblock \bibinfo{journal}{\emph{arXiv preprint arXiv:2307.09288}} (\bibinfo{year}{2023}).
\newblock


\bibitem[Van~Balen and Levy(2019)]%
        {van2019pq}
\bibfield{author}{\bibinfo{person}{Jan Van~Balen} {and} \bibinfo{person}{Mark Levy}.} \bibinfo{year}{2019}\natexlab{}.
\newblock \showarticletitle{PQ-VAE: Efficient Recommendation Using Quantized Embeddings.}. In \bibinfo{booktitle}{\emph{RecSys (Late-Breaking Results)}}. \bibinfo{pages}{46--50}.
\newblock


\bibitem[Van Den~Oord et~al\mbox{.}(2017)]%
        {vq-vae}
\bibfield{author}{\bibinfo{person}{Aaron Van Den~Oord}, \bibinfo{person}{Oriol Vinyals}, {et~al\mbox{.}}} \bibinfo{year}{2017}\natexlab{}.
\newblock \showarticletitle{Neural discrete representation learning}.
\newblock \bibinfo{journal}{\emph{Advances in neural information processing systems}}  \bibinfo{volume}{30} (\bibinfo{year}{2017}).
\newblock


\bibitem[Vaswani et~al\mbox{.}(2017)]%
        {vaswani2017attention}
\bibfield{author}{\bibinfo{person}{Ashish Vaswani}, \bibinfo{person}{Noam Shazeer}, \bibinfo{person}{Niki Parmar}, \bibinfo{person}{Jakob Uszkoreit}, \bibinfo{person}{Llion Jones}, \bibinfo{person}{Aidan~N Gomez}, \bibinfo{person}{{\L}ukasz Kaiser}, {and} \bibinfo{person}{Illia Polosukhin}.} \bibinfo{year}{2017}\natexlab{}.
\newblock \showarticletitle{Attention is all you need}.
\newblock \bibinfo{journal}{\emph{Advances in neural information processing systems}}  \bibinfo{volume}{30} (\bibinfo{year}{2017}).
\newblock


\bibitem[Wang et~al\mbox{.}(2020)]%
        {linformer}
\bibfield{author}{\bibinfo{person}{Sinong Wang}, \bibinfo{person}{Belinda~Z Li}, \bibinfo{person}{Madian Khabsa}, \bibinfo{person}{Han Fang}, {and} \bibinfo{person}{Hao Ma}.} \bibinfo{year}{2020}\natexlab{}.
\newblock \showarticletitle{Linformer: Self-attention with linear complexity}.
\newblock \bibinfo{journal}{\emph{arXiv preprint arXiv:2006.04768}} (\bibinfo{year}{2020}).
\newblock


\bibitem[Wang et~al\mbox{.}(2022)]%
        {nci}
\bibfield{author}{\bibinfo{person}{Yujing Wang}, \bibinfo{person}{Yingyan Hou}, \bibinfo{person}{Haonan Wang}, \bibinfo{person}{Ziming Miao}, \bibinfo{person}{Shibin Wu}, \bibinfo{person}{Qi Chen}, \bibinfo{person}{Yuqing Xia}, \bibinfo{person}{Chengmin Chi}, \bibinfo{person}{Guoshuai Zhao}, \bibinfo{person}{Zheng Liu}, {et~al\mbox{.}}} \bibinfo{year}{2022}\natexlab{}.
\newblock \showarticletitle{A neural corpus indexer for document retrieval}.
\newblock \bibinfo{journal}{\emph{Advances in Neural Information Processing Systems}}  \bibinfo{volume}{35} (\bibinfo{year}{2022}), \bibinfo{pages}{25600--25614}.
\newblock


\bibitem[Wu et~al\mbox{.}(2023)]%
        {wu2023hyperparameter}
\bibfield{author}{\bibinfo{person}{Di Wu}, \bibinfo{person}{Bo Sun}, {and} \bibinfo{person}{Mingsheng Shang}.} \bibinfo{year}{2023}\natexlab{}.
\newblock \showarticletitle{Hyperparameter learning for deep learning-based recommender systems}.
\newblock \bibinfo{journal}{\emph{IEEE Transactions on Services Computing}} (\bibinfo{year}{2023}).
\newblock


\bibitem[Wu et~al\mbox{.}(2021)]%
        {wu2021linear}
\bibfield{author}{\bibinfo{person}{Yongji Wu}, \bibinfo{person}{Defu Lian}, \bibinfo{person}{Neil~Zhenqiang Gong}, \bibinfo{person}{Lu Yin}, \bibinfo{person}{Mingyang Yin}, \bibinfo{person}{Jingren Zhou}, {and} \bibinfo{person}{Hongxia Yang}.} \bibinfo{year}{2021}\natexlab{}.
\newblock \showarticletitle{Linear-time self attention with codeword histogram for efficient recommendation}. In \bibinfo{booktitle}{\emph{Proceedings of the Web Conference 2021}}. \bibinfo{pages}{1262--1273}.
\newblock


\bibitem[Xia et~al\mbox{.}(2024)]%
        {xia2024achieving}
\bibfield{author}{\bibinfo{person}{Yan Xia}, \bibinfo{person}{Hai Huang}, \bibinfo{person}{Jieming Zhu}, {and} \bibinfo{person}{Zhou Zhao}.} \bibinfo{year}{2024}\natexlab{}.
\newblock \showarticletitle{Achieving Cross Modal Generalization with Multimodal Unified Representation}.
\newblock \bibinfo{journal}{\emph{Advances in Neural Information Processing Systems}}  \bibinfo{volume}{36} (\bibinfo{year}{2024}).
\newblock


\bibitem[Yang et~al\mbox{.}(2023)]%
        {yang2023vqgraph}
\bibfield{author}{\bibinfo{person}{Ling Yang}, \bibinfo{person}{Ye Tian}, \bibinfo{person}{Minkai Xu}, \bibinfo{person}{Zhongyi Liu}, \bibinfo{person}{Shenda Hong}, \bibinfo{person}{Wei Qu}, \bibinfo{person}{Wentao Zhang}, \bibinfo{person}{Bin Cui}, \bibinfo{person}{Muhan Zhang}, {and} \bibinfo{person}{Jure Leskovec}.} \bibinfo{year}{2023}\natexlab{}.
\newblock \showarticletitle{Vqgraph: Graph vector-quantization for bridging gnns and mlps}.
\newblock \bibinfo{journal}{\emph{arXiv preprint arXiv:2308.02117}} (\bibinfo{year}{2023}).
\newblock


\bibitem[Yue et~al\mbox{.}(2023)]%
        {yue2023llamarec}
\bibfield{author}{\bibinfo{person}{Zhenrui Yue}, \bibinfo{person}{Sara Rabhi}, \bibinfo{person}{Gabriel de Souza~Pereira Moreira}, \bibinfo{person}{Dong Wang}, {and} \bibinfo{person}{Even Oldridge}.} \bibinfo{year}{2023}\natexlab{}.
\newblock \showarticletitle{LlamaRec: Two-stage recommendation using large language models for ranking}.
\newblock \bibinfo{journal}{\emph{arXiv preprint arXiv:2311.02089}} (\bibinfo{year}{2023}).
\newblock


\bibitem[Zeghidour et~al\mbox{.}(2021)]%
        {zeghidour2021soundstream}
\bibfield{author}{\bibinfo{person}{Neil Zeghidour}, \bibinfo{person}{Alejandro Luebs}, \bibinfo{person}{Ahmed Omran}, \bibinfo{person}{Jan Skoglund}, {and} \bibinfo{person}{Marco Tagliasacchi}.} \bibinfo{year}{2021}\natexlab{}.
\newblock \showarticletitle{Soundstream: An end-to-end neural audio codec}.
\newblock \bibinfo{journal}{\emph{IEEE/ACM Transactions on Audio, Speech, and Language Processing}}  \bibinfo{volume}{30} (\bibinfo{year}{2021}), \bibinfo{pages}{495--507}.
\newblock


\bibitem[Zhan et~al\mbox{.}(2024)]%
        {zhan2024anygpt}
\bibfield{author}{\bibinfo{person}{Jun Zhan}, \bibinfo{person}{Junqi Dai}, \bibinfo{person}{Jiasheng Ye}, \bibinfo{person}{Yunhua Zhou}, \bibinfo{person}{Dong Zhang}, \bibinfo{person}{Zhigeng Liu}, \bibinfo{person}{Xin Zhang}, \bibinfo{person}{Ruibin Yuan}, \bibinfo{person}{Ge Zhang}, \bibinfo{person}{Linyang Li}, {et~al\mbox{.}}} \bibinfo{year}{2024}\natexlab{}.
\newblock \showarticletitle{AnyGPT: Unified Multimodal LLM with Discrete Sequence Modeling}.
\newblock \bibinfo{journal}{\emph{arXiv preprint arXiv:2402.12226}} (\bibinfo{year}{2024}).
\newblock


\bibitem[Zhang et~al\mbox{.}(2023a)]%
        {zhang2023query}
\bibfield{author}{\bibinfo{person}{Jin Zhang}, \bibinfo{person}{Defu Lian}, \bibinfo{person}{Haodi Zhang}, \bibinfo{person}{Baoyun Wang}, {and} \bibinfo{person}{Enhong Chen}.} \bibinfo{year}{2023}\natexlab{a}.
\newblock \showarticletitle{Query-Aware Quantization for Maximum Inner Product Search}. In \bibinfo{booktitle}{\emph{Proceedings of the AAAI Conference on Artificial Intelligence}}, Vol.~\bibinfo{volume}{37}. \bibinfo{pages}{4875--4883}.
\newblock


\bibitem[Zhang et~al\mbox{.}(2023b)]%
        {zhang2023regularized}
\bibfield{author}{\bibinfo{person}{Jiahui Zhang}, \bibinfo{person}{Fangneng Zhan}, \bibinfo{person}{Christian Theobalt}, {and} \bibinfo{person}{Shijian Lu}.} \bibinfo{year}{2023}\natexlab{b}.
\newblock \showarticletitle{Regularized vector quantization for tokenized image synthesis}. In \bibinfo{booktitle}{\emph{Proceedings of the IEEE/CVF Conference on Computer Vision and Pattern Recognition}}. \bibinfo{pages}{18467--18476}.
\newblock


\bibitem[Zhang et~al\mbox{.}(2018)]%
        {zhang2018efficient}
\bibfield{author}{\bibinfo{person}{Kunpeng Zhang}, \bibinfo{person}{Shaokun Fan}, {and} \bibinfo{person}{Harry~Jiannan Wang}.} \bibinfo{year}{2018}\natexlab{}.
\newblock \showarticletitle{An efficient recommender system using locality sensitive hashing}.
\newblock  (\bibinfo{year}{2018}).
\newblock


\bibitem[Zhao et~al\mbox{.}(2021)]%
        {zhao2021embedding}
\bibfield{author}{\bibinfo{person}{Jing Zhao}, \bibinfo{person}{Jingya Wang}, \bibinfo{person}{Madhav Sigdel}, \bibinfo{person}{Bopeng Zhang}, \bibinfo{person}{Phuong Hoang}, \bibinfo{person}{Mengshu Liu}, {and} \bibinfo{person}{Mohammed Korayem}.} \bibinfo{year}{2021}\natexlab{}.
\newblock \showarticletitle{Embedding-based recommender system for job to candidate matching on scale}.
\newblock \bibinfo{journal}{\emph{arXiv preprint arXiv:2107.00221}} (\bibinfo{year}{2021}).
\newblock


\bibitem[Zheng et~al\mbox{.}(2023)]%
        {zheng2023adapting}
\bibfield{author}{\bibinfo{person}{Bowen Zheng}, \bibinfo{person}{Yupeng Hou}, \bibinfo{person}{Hongyu Lu}, \bibinfo{person}{Yu Chen}, \bibinfo{person}{Wayne~Xin Zhao}, {and} \bibinfo{person}{Ji-Rong Wen}.} \bibinfo{year}{2023}\natexlab{}.
\newblock \showarticletitle{Adapting Large Language Models by Integrating Collaborative Semantics for Recommendation}.
\newblock \bibinfo{journal}{\emph{arXiv preprint arXiv:2311.09049}} (\bibinfo{year}{2023}).
\newblock


\bibitem[Zheng et~al\mbox{.}(2024)]%
        {zheng2024unicode}
\bibfield{author}{\bibinfo{person}{Sipeng Zheng}, \bibinfo{person}{Bohan Zhou}, \bibinfo{person}{Yicheng Feng}, \bibinfo{person}{Ye Wang}, {and} \bibinfo{person}{Zongqing Lu}.} \bibinfo{year}{2024}\natexlab{}.
\newblock \showarticletitle{UniCode: Learning a Unified Codebook for Multimodal Large Language Models}.
\newblock \bibinfo{journal}{\emph{arXiv preprint arXiv:2403.09072}} (\bibinfo{year}{2024}).
\newblock


\end{thebibliography}
